\documentclass{dynacontalk}
\setcopyright{none}

\AtBeginDocument{}
\usepackage{amsmath}
\usepackage{graphicx}
\usepackage{multirow}
\usepackage{booktabs}
\usepackage{array}
\usepackage{tabularx}
\usepackage{placeins}

\begin{document}

\title[DynaConTalk]{DynaConTalk: Wavelet-Constrained Diffusion for Long-Form and Controllable Holistic Co-Speech 3D Motion}

\author{Yifei Zhu}
\orcid{0009-0002-8895-3553}
\affiliation{%
  \institution{Tohoku University}
  \city{Sendai}
  \country{Japan}}
\email{zhuyifei.q4@dc.tohoku.ac.jp}

\author{Yangyang Cai}
\orcid{0009-0009-2605-8511}
\affiliation{%
  \institution{Tohoku University}
  \city{Sendai}
  \country{Japan}}

\author{Mingyi Shi}
\orcid{0000-0002-5180-600X}
\affiliation{%
  \institution{University of Hong Kong}
  \city{Hong Kong}
  \country{Hong Kong}}

\author{Miao Cheng}
\orcid{0000-0003-0611-6170}
\affiliation{%
  \institution{Tohoku University}
  \city{Sendai}
  \country{Japan}}

\author{Lin Gu}
\orcid{0000-0002-7504-031X}
\affiliation{%
  \institution{Tohoku University}
  \city{Sendai}
  \country{Japan}}

\author{Taku Komura}
\authornote{Corresponding authors.}
\orcid{0000-0002-2729-5860}
\affiliation{%
  \institution{University of Hong Kong}
  \city{Hong Kong}
  \country{Hong Kong}}
\email{taku@cs.hku.hk}

\author{Yoshifumi Kitamura}
\authornotemark[1]
\orcid{0000-0002-7047-627X}
\affiliation{%
  \institution{Tohoku University}
  \city{Sendai}
  \country{Japan}}
\email{kitamura@riec.tohoku.ac.jp}

\renewcommand{\shortauthors}{Zhu et al.}

\begin{abstract}
Holistic co-speech animation is prone to averaging in both motion representation
and speech conditioning. In coordinate-space diffusion, slow body posture,
mid-frequency gesture strokes, and fast hand or facial details are entangled in
one prediction target, often producing low-variance, over-smoothed motion.
Meanwhile, dense rhythmic and acoustic cues can dominate sparse
content-specific information under fixed multimodal fusion. We present
DynaConTalk, a wavelet-constrained diffusion framework for long-form and
controllable holistic co-speech motion generation. Diffusion operates in
stationary wavelet transform (SWT) coefficient space, whose temporally aligned
bands separate coarse posture evolution, gesture strokes, and fine expressive
details. Our dynamic gating network preserves HuBERT and speaker identity as
a base and selectively adds rhythm, mel, and transcript features through
motion-state- and noise-aware residual gates. Attention pooling and learned
depth routing deliver complementary conditions to each denoising stage, while
a frame-resolution rhythm path preserves precise timing. A signed
proposal--consensus update then reconciles these conditions with the evolving
motion state. Matched-noise constraint injection uses the same sampling
interface for history continuation and localized keypose repair, and extends
to reference-guided control. Separate body-hand and facial denoisers, followed
by inverse SWT and a pose-driven root regressor, produce holistic motion.
Experiments evaluate generation quality, facial accuracy, temporal continuity,
and controllable editing. Code, models, and the interactive editing interface
are available at
\url{https://github.com/zhuyifeiabcd1/DynaConTalk}.

\end{abstract}


\begin{teaserfigure}
  \centering
  \includegraphics[width=\textwidth]{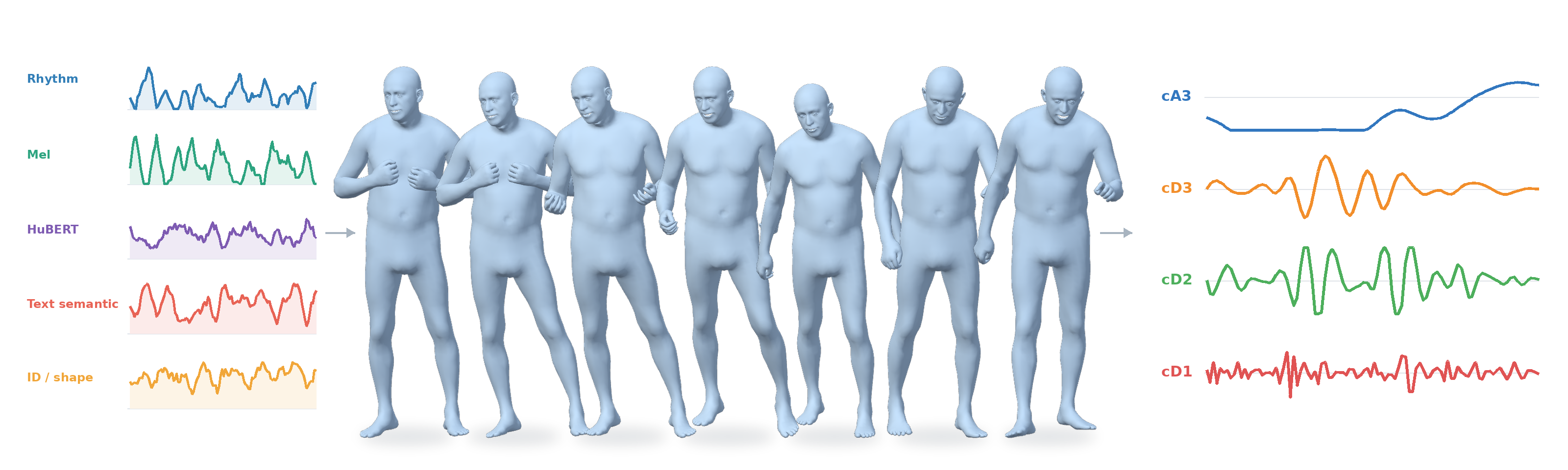}
  \caption{\textbf{DynaConTalk maps multimodal speech cues to holistic
  3D co-speech motion in an aligned wavelet space.}
  The SWT bands separate coarse body evolution, gesture-scale dynamics,
  and fine expressive details while preserving temporal alignment.}
  \Description{Multimodal speech conditions on the left drive a sequence
  of holistic human poses in the center. Four aligned SWT bands on the
  right represent coarse-to-fine motion dynamics.}
  \label{fig:teaser}
\end{teaserfigure}

\maketitle
\hypersetup{pdfcreator={LaTeX},pdfsubject={Holistic co-speech motion generation}}
\fancypagestyle{arxivpage}{%
  \fancyhf{}%
  \renewcommand{\headrulewidth}{0pt}%
  \renewcommand{\footrulewidth}{0pt}%
  \fancyhead[LE]{\sffamily\footnotesize\thepage\quad\textbullet\quad Zhu et al.}%
  \fancyhead[RO]{\sffamily\footnotesize DynaConTalk\quad\textbullet\quad\thepage}%
}
\fancypagestyle{firstpagestyle}{%
  \fancyhf{}%
  \renewcommand{\headrulewidth}{0pt}%
  \renewcommand{\footrulewidth}{0pt}%
  \fancyfoot[C]{\sffamily\footnotesize\thepage}%
}
\pagestyle{arxivpage}
\thispagestyle{firstpagestyle}

\section{Introduction}
Holistic co-speech generation is inherently prone to averaging, largely because the motion signal itself is layered rather than a single homogeneous pose sequence.
While the torso evolves slowly to anchor the speaker, arms and hands
produce mid-frequency gesture strokes, and fingers and facial expressions carry
fast expressive details~\cite{alexanderson2023listen,ao2023gesturediffuclip,yang2023diffusestylegesture,yi2023generating,liu2024emage,chen2024diffsheg,pavlakos2019expressive,zhang2025semtalk}. This layered structure makes holistic co-speech
generation especially vulnerable to averaging. When all body, hand, face, and
trajectory variables are diffused as one coordinate vector, slow posture,
gesture strokes, and high-frequency details are optimized in the same space.
The model can then reduce loss by producing safe and over-smoothed motion,
where rare expressive gestures are absorbed into common beat-like movements.

A second source of averaging is rooted in the way speech conditioning is structured.
Speech provides multiple heterogeneous cues: rhythm and acoustic energy are dense and frequent,
while semantic and textual cues are sparse but crucial for content-specific gestures.
Fixed concatenation or static fusion tends to let frequent rhythmic cues dominate
the generated update, causing rare semantic gestures to be weakened or replaced
by generic motion.

To cope with these problems, we propose DynaConTalk, a wavelet-constrained diffusion framework for long-form and controllable holistic co-speech motion generation. Our approach tackles both sources of averaging by explicitly structuring the motion representation and conditioning signals according to their temporal and semantic properties.

{\it First}, to address motion-level averaging, we perform diffusion in the stationary wavelet transform (SWT) space, where aligned bands separate coarse posture, mid-frequency gestures, and fine expressive details. This decomposition enables scale-specific learning through dedicated prediction heads, preventing low-amplitude expressive dynamics from being overwhelmed by dominant coarse motion without resorting to a single entangled coordinate-space target.

{\it Second}, we introduce state-aware dynamic gating and
proposal--consensus conditioning. HuBERT features and speaker identity form a
persistent base; independent residual gates select rhythm, mel, and transcript
cues using both the current noisy motion and the diffusion timestep. Learned
attention pooling and modality-to-depth routing adapt these signals to each
denoising stage, while a separate frame-level rhythm path preserves timing
within temporal patches. The consensus branch controls a signed motion update,
allowing a proposed change to be accepted, attenuated, or reversed according to
the current motion state.

{\it Third}, we introduce matched-noise constraint injection for
long-form continuation and flexible user control. A clean motion history or
reference target is re-noised to the current diffusion timestep before being
injected into the constrained region. This provides a shared sampling
interface for window continuation, localized keypose repair, and
reference-guided control. The released window sampler also uses a short
velocity-based blend at joins, and the interactive editor repairs only clips
surrounding the requested edits.

Our contributions are summarized as follows:
\begin{itemize}
  \item \textbf{SWT coefficient diffusion.}
  We formulate motion generation in temporally aligned wavelet
  space, decoupling coarse posture, gesture strokes, and fine hand and facial
  details through band-specific decoding.

  \item \textbf{Dynamic conditioning.}
  We combine a persistent base, sparse residual gates, attention pooling,
  depth routing, and a frame-level rhythm path with signed proposal--consensus
  injection into the denoiser.

  \item \textbf{Matched-noise constraint injection.}
  We turn window continuation, keypose editing, and reference control into a
  single in-distribution denoising process by injecting constraints at the
  matched diffusion noise level.
\end{itemize}

\section{Related Work}

\subsection{From Neural Regression to Holistic Diffusion}

Early speech-driven animation mapped linguistic and prosodic cues to authored
gesture libraries~\cite{cassell2001beat,kopp2006towards,prendinger2004mpml,kipp2005gesture}.
Neural methods replaced these discrete pipelines with direct speech-to-motion
regression using convolutional and recurrent models
~\cite{ginosar2019learning,yoon2019robots,hasegawa2018evaluation,
ferstl2018investigating,takeuchi2017speech}, but deterministic regression tends
to average over the one-to-many mapping between speech and valid gestures
~\cite{cassell1999speech,loehr2007aspects,kucherenko2021moving}.

Generative models address this ambiguity through adversarial learning
~\cite{ferstl2020adversarial,bhattacharya2021speech2affectivegestures,
wu2021modeling}, variational models~\cite{li2021audio2gestures,petrovich2021action,Pang_2023},
and normalizing flows~\cite{henter2020moglow,alexanderson2020style,ye2022audio}.
Diffusion models~\cite{ddpm,ddim} further improved stochastic motion generation
~\cite{mdm,mld} and speech-driven gesture synthesis
~\cite{alexanderson2023listen,zhu2023taming,ao2023gesturediffuclip,
yang2023diffusestylegesture}. Recent systems have expanded from upper-body
gesture to coordinated body, hands, face, and trajectory
~\cite{ng2021body2hands,Pang_2023,yi2023generating,liu2024emage,
chen2024diffsheg,liu2024towards,ng2024audio,xu2024chain,zhang2025semtalk}.
Most generate motion in pose-coordinate or learned latent spaces; DynaConTalk
instead uses an explicit temporally aligned multiscale representation as the
diffusion target.

\subsection{Content, Rhythm, and Dynamic Speech Conditioning}

Co-speech motion depends on complementary speech cues: prosody and acoustics
capture rhythm and emphasis, self-supervised features such as wav2vec and
HuBERT encode phonetic and contextual information
~\cite{baevski2020wav2vec,hsu2021hubert}, while text and speaker identity
provide semantic and behavioral context. Multimodal methods combine these
signals or learn cross-modal associations
~\cite{liu2022learning,liang2022seeg,zhi2023livelyspeaker}, and DisCo separates
content and rhythm representations for diverse gesture generation
~\cite{liu2022disco}.

Semantic conditioning has also been introduced through text features
~\cite{kucherenko2020gesticulator}, language-aligned diffusion latents
~\cite{ao2023gesturediffuclip}, retrieved gesture concepts
~\cite{Zhang_2024}, and holistic semantic-motion modeling
~\cite{zhang2025semtalk}. DynaConTalk instead dynamically regulates multiple
speech conditions according to the evolving motion state.

\subsection{Frequency and Multiscale Motion Representation}

Frequency structure has been used to model periodic and long-range motion
~\cite{holden2017phase,starke2022deepphase}. In gesture synthesis, DisCo uses
motion analysis to disentangle content and rhythm without generating directly
in the frequency domain~\cite{liu2022disco}. More recent work incorporates
Fourier or wavelet structure into motion modeling
~\cite{feng2024motionwavelethumanmotionprediction,
fan2025zerozeroshotmotiongeneration}, while wavelet-domain diffusion has also
been explored for image synthesis and restoration
~\cite{phung2023waveletdiffusionmodelsfast,
huang2024wavedmwaveletbaseddiffusionmodels}.

Co-speech motion additionally requires temporal alignment because speech cues,
gesture events, continuation boundaries, and user controls occur at specific
times. Unlike global Fourier representations or decimated wavelets, the
stationary wavelet transform (SWT) retains coefficients at every frame and
scale. Following diffusion inpainting methods such as RePaint
~\cite{Lugmayr_2022_CVPR}, known content can be re-noised to the current
diffusion level and reintroduced during reverse sampling. DynaConTalk exploits this alignment for both speech-conditioned
generation and localized control. This temporal alignment also provides a common interface for long-form synthesis and motion control: whereas prior work uses overlapping diffusion windows or inpainting/repainting for continuation and separate style, semantic, joint, or keyframe constraints for control~\cite{mdm,jin2023actwishfinegrainedcontrol,chen2024tamingdiffusionprobabilisticmodels,ahuja2020style,chen2024enabling,Zhang_2024}, DynaConTalk represents history, keyposes, and reference motion as partial observations injected at matched diffusion noise levels.

\section{Method}
\begin{figure*}[t]
    \centering
    \includegraphics[width=\textwidth]{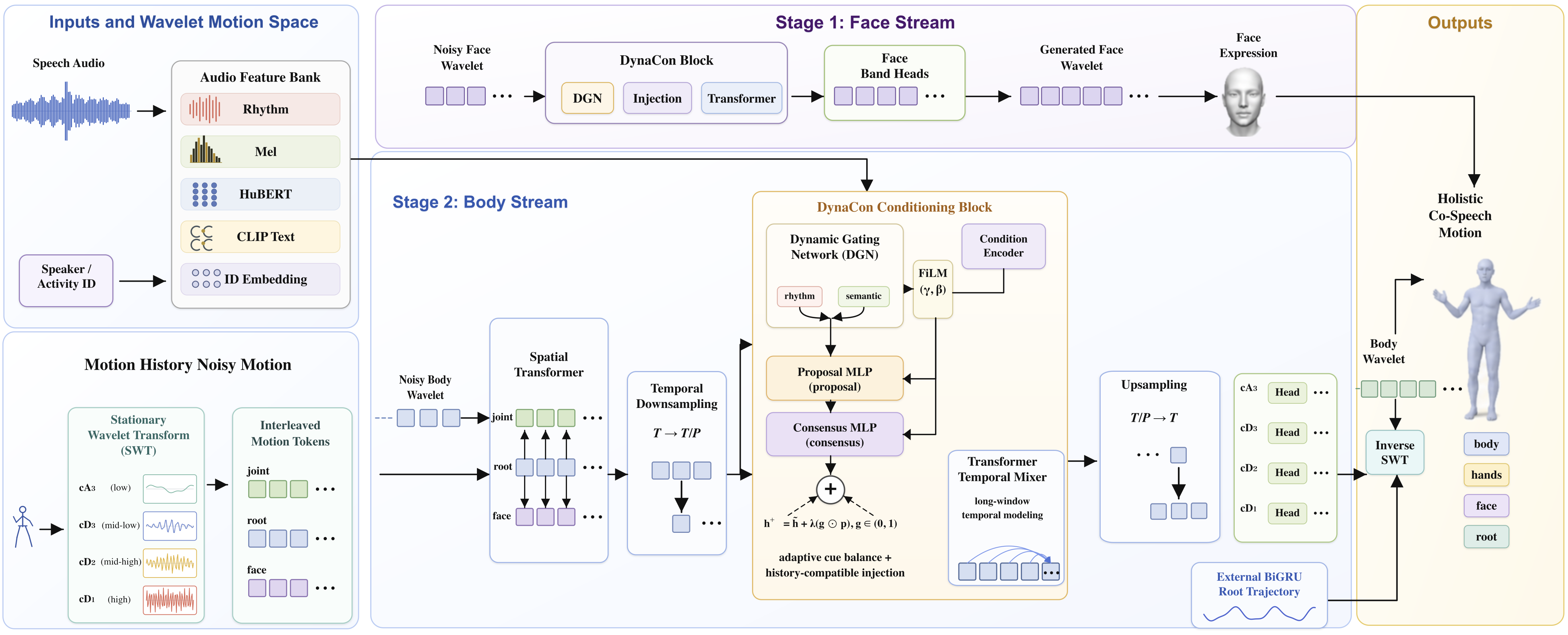}
    \caption{
    Overview of DynaConTalk. Speech conditions drive separate body-hand and
    face wavelet diffusion streams. The body stream uses structural motion
    tokens, the face stream predicts expression coefficients, and a
    pose-driven BiGRU supplies root translation after local motion generation.
    Matched-noise constraints support window continuation and keypose editing.
    }
    \Description{The DynaConTalk architecture diagram showing multimodal
    speech conditioning, body and face wavelet diffusion streams, root
    trajectory regression, and matched-noise control.}
    \label{fig:framework}
\end{figure*}

Co-speech animation must turn rapidly varying speech into physically coherent
body, hand, face, and root motion over long durations. DynaConTalk addresses this
by performing conditional diffusion in a stationary wavelet motion space, where
multi-scale motion coefficients are generated by stream-specific denoisers and
reconstructed into SMPL-X~\cite{pavlakos2019expressive} motion by inverse SWT.
We first define the motion representation and wavelet diffusion
(Secs.~3.1 and~3.2), then describe the structure-aware denoisers (Sec.~3.3),
state-aware dynamic gating and consensus injection (Sec.~3.4), matched-noise
continuation and control (Sec.~3.5), and the training objectives (Sec.~3.6).

\subsection{Problem Formulation}

Given an arbitrary-length speech signal, we generate holistic co-speech motion
including body-hand pose, facial expression, and root trajectory. Training
sequences are cut into $N$-frame clips. For each clip, the audio is encoded into
a frame-aligned feature bank containing a three-channel rhythm descriptor
(amplitude envelope, short-time energy, and onsets), a 128-band mel
spectrogram, and 1024-dimensional HuBERT-Large features. A 512-dimensional
CLIP transcript embedding provides window-level text context. Speaker identity
is represented by the speaker index, a learned speaker embedding, and the
first ten SMPL-X shape coefficients. The local motion consists of body-hand
6D rotations $B\in\mathbb{R}^{N\times330}$ and facial expression coefficients
$F\in\mathbb{R}^{N\times100}$.

We model body-hand and face motion in SWT coefficient space. With a 3-level SWT,
each channel is represented by four aligned bands
$(\mathrm{cA}_3,\mathrm{cD}_3,\mathrm{cD}_2,\mathrm{cD}_1)$, giving body-hand
targets $X^B_0\in\mathbb{R}^{N\times1320}$ and face targets
$X^F_0\in\mathbb{R}^{N\times400}$. We find that diffusing the root trajectory
jointly with the body motion is ineffective for two reasons. First, audio
is largely uninformative for root trajectory prediction: the root trajectory is
mainly correlated with lower-body motion. Joint training may
produce plausible trajectories conditioned on the audio, but these trajectories
are not necessarily consistent with the generated motion. Second, the root trajectory
is only 3-dimensional, whereas the body representation is 330-dimensional,
making its loss easily underweighted during training despite its strong visual
impact. We therefore attach a separate pose-driven root-translation regressor.
It uses pelvis-centered joint positions, their temporal derivatives, and root
yaw rate extracted from the generated body pose. A convolutional patch
encoder, bidirectional GRU, and temporal upsampling decoder predict translation
relative to the first frame. Root translation is added after local pose
generation; it is not part of the body denoiser's 1320-dimensional target.

\subsection{Wavelet-Domain Diffusion}
\label{waveletdiff}
As shown in the left part of Fig.~\ref{fig:framework}, DynaConTalk maps local
motion into a stationary wavelet space before diffusion. For a stream target
$M\in\{B,F\}$, we apply SWT along the temporal axis:
\[
    X_0=\mathcal{W}(M)
    =\{\mathrm{cA}_3,\mathrm{cD}_3,\mathrm{cD}_2,\mathrm{cD}_1\}.
\]
The coefficients are interleaved per channel and normalized by training-set
mean and standard deviation. This representation exposes slow posture evolution
and fast expressive details as temporally aligned bands, allowing the denoiser
to model coarse motion and fine details in a shared generation space.

The forward diffusion process corrupts the normalized wavelet target $x_0$ by
\[
    q(x_t|x_0)=
    \mathcal{N}\left(x_t;\sqrt{\bar{\alpha}_t}x_0,
    (1-\bar{\alpha}_t)\mathbf{I}\right).
\]
Given $x_t$, timestep $t$, and condition $c$, the denoiser predicts the clean
wavelet sample:
\[
    \hat{x}_0=f_\theta(x_t,t,c).
\]
After sampling, the predicted coefficients are de-normalized and reconstructed
by inverse SWT. This frequency-aware prediction makes isolated coordinate errors
less likely to appear as abrupt frame-wise motion artifacts after reconstruction.
We provide additional diagnostic analysis of the SWT motion space in
Fig.~\ref{fig:band_structure}.

\subsection{Structure-Aware Denoiser}
The  body-hand stream and the face stream
share the same diffusion formulation and conditioning interface,
but use stream-specific backbones. The body-hand stream predicts high-dimensional
SMPL-X joint rotations and uses structural motion tokens, while the
face stream predicts compact expression coefficients and uses a lighter temporal
Transformer without joint tokenization.

\paragraph{Joint- and band-aware motion encoder.}
The body target is reshaped to $N\times55\times4\times6$: each of the 55
SMPL-X joints has a six-dimensional rotation in each of four aligned SWT
bands. A band-specific temporal convolution embeds each joint-band sequence.
Learned joint and band embeddings identify the tokens. Within each frame,
a spatial Transformer shared across bands models inter-joint coordination;
a second Transformer mixes the four bands of each joint. Band fusion followed
by joint fusion produces a frame-level feature. A linear projection of the
input provides a parallel bypass:
\[
 H=\mathrm{Linear}(x_t)
   +\rho\,\mathrm{LN}\!\left(\mathrm{TokenEncoder}(x_t)\right),
\]
where $\rho$ is initialized to zero. This allows training to start with the
linear representation and progressively incorporate the structured token
branch. The face stream uses a direct linear projection instead of joint
tokenization.

\paragraph{Local and patch-level temporal processing.}
Each denoiser has four stages. A stage combines two local convolutional
modules with a patch-level temporal mixer. Pointwise and depthwise temporal
convolutions preserve local detail. The mixer downsamples motion features by
$8\times$, applies stage-specific FiLM and proposal--consensus conditioning,
and processes the resulting patches with a temporal Transformer. A reversed
copy of the patch sequence is prepended before attention, and the outputs for
the original sequence are retained. Patch features are repeated back to
frame resolution and fused with the frame-level branch. A diffusion-time
token accompanies the sequence; the identity projection is added to this
token in both classifier-free guidance branches. Thus, local motion detail,
longer-range temporal context, and speaker identity are available throughout
the trunk.

\paragraph{Four band-specific wavelet decoding heads.}
The SWT representation also affects the output decoding design. Instead of using one shared decoder for all coefficients, DynaConTalk predicts each wavelet band with an independent head:
\[
    \hat{x}_0^{b}=\mathrm{Head}_{b}(h),
    \qquad
    b\in\{\mathrm{cA}_3,\mathrm{cD}_3,\mathrm{cD}_2,\mathrm{cD}_1\}.
\]
The low-frequency head $\mathrm{cA}_3$ models stable posture and broad body
evolution, $\mathrm{cD}_3$ and $\mathrm{cD}_2$ capture gesture strokes and
rhythmic mid-frequency motion, and $\mathrm{cD}_1$ preserves fast hand and facial
details. This is a direct benefit of the wavelet motion space: different
temporal scales no longer compete inside a single output decoder. The model can
allocate decoder capacity according to frequency-specific motion statistics,
which reduces over-smoothing of high-frequency details while keeping the inverse
SWT reconstruction coherent.

\subsection{State-Aware Dynamic Gating and Consensus Conditioning}
\label{sec:dgn}

Dense acoustic cues and sparse content cues need not contribute equally to
every frame or every denoising stage. Our Dynamic Gating Network (DGN)
separates a persistent semantic--identity base from selectively activated
residual streams. Its residual gates depend on speech, the encoded noisy
motion, and the diffusion noise level. We use $i$ for a motion frame, $s$
for a diffusion timestep, and $\ell$ for a denoiser stage.

\paragraph{Persistent base and residual modalities.}
Separate MLPs project rhythm, mel, HuBERT, and CLIP text features into a common
embedding space. Let $e_i^{\mathrm{hub}}$, $e_i^{\mathrm{rh}}$,
$e_i^{\mathrm{mel}}$, and $e_i^{\mathrm{txt}}$ denote these embeddings.
The window transcript embedding is broadcast across its frames; it is not a
per-word label. An identity MLP embeds the normalized speaker index and shape
coefficients, and a learned speaker lookup embedding is added to it to form
$e^{\mathrm{id}}$. The base is
\[
 b_i=e_i^{\mathrm{hub}}+e^{\mathrm{id}}.
\]
HuBERT supplies frame-aligned speech context, while identity anchors
speaker-specific behavior. The residual modality set is
$\mathcal{M}=\{\mathrm{rh},\mathrm{mel},\mathrm{txt}\}$. These streams have
independent sigmoid gates rather than competing through a modality softmax.
Consequently, several cues can be active together, or all residual cues can
be attenuated while the base remains available.

\paragraph{Motion-state- and noise-aware gates.}
Let $H_i^{(s)}$ be the frame-level encoding of the noisy wavelet input
before the denoiser trunk, and let $v_s$ be the diffusion-time embedding.
Their projected sum gives a gate context:
\[
 q_{i,s}=\mathrm{LN}\!\left(W_hH_i^{(s)}+W_sv_s\right).
\]
The gate MLP receives normalized residual embeddings and this context:
\[
 z_{i,s}=
 G\!\left([\mathrm{LN}(e_i^{\mathrm{rh}});
          \mathrm{LN}(e_i^{\mathrm{mel}});
          \mathrm{LN}(e_i^{\mathrm{txt}});q_{i,s}]\right).
\]
For residual modality $m$, the gate is
\begin{equation}
\label{eq:dgn_gate}
 \alpha_{i,s}^{m}
 =\operatorname{sigmoid}\!\left(
 z_{i,s}^{m}+\eta_m\left(\frac{s}{S}-\frac12\right)\right),
\end{equation}
where $S=1000$ is the training diffusion schedule length and $\eta_m$ is
a learned noise-level coefficient. The added linear term permits
modality-specific changes in gate opening across the denoising schedule;
the state-aware MLP can also adapt to the current sample. Normalization is
used in the gate inputs, while the residual values retain their learned
amplitudes. Define $u_i^0=b_i$ and
$u_i^m=\alpha_{i,s}^m e_i^m$ for $m\in\mathcal{M}$. The semantic base and
these three residual streams are retained separately for stage routing.

The final gate layer has zero-initialized weights and bias $-3$, so the
residual gates start at $\operatorname{sigmoid}(-3)\simeq0.047$.
The noise coefficients $\eta_m$ start at zero. A small mean-gate penalty
(Sec.~3.6) encourages selective residual use; it does not impose hard
binary decisions.

\paragraph{Attention pooling and modality-to-depth routing.}
To align conditions with the motion trunk's eight-frame patches, each source
stream is pooled independently with a shared learned score function $a$.
For source $k$ and patch $\mathcal{P}_p$,
\[
 \omega_{i,p}^{k}
 =\frac{\exp a(u_i^k)}
        {\sum_{j\in\mathcal{P}_p}\exp a(u_j^k)},\qquad
 \bar{u}_p^k=\sum_{i\in\mathcal{P}_p}\omega_{i,p}^{k}u_i^k.
\]
The score function's last layer starts at zero, giving uniform averaging
initially. Learning can then emphasize relevant frames within a patch.
Each stage has its own learned routing logits $r_{\ell,k}$ over the four
source streams:
\[
 w_{\ell,k}=\operatorname{softmax}_{k}(r_{\ell,k}),\qquad
 c_{\ell,p}=\mathcal{C}\!\left(\sum_k w_{\ell,k}\bar{u}_p^k\right).
\]
Here, $\mathcal{C}$ uses pointwise channel projections, layer normalization,
and dropout. Routing weights are learned per stage and shared across samples;
the residual gates and attention-pooling weights vary with the input.
Zero-initialized routing logits initially allocate equal weights. Separate
stage projections turn $c_{\ell,p}$ into a proposal feature $f_{\ell,p}$
and FiLM parameters $(\gamma_{\ell,p},\beta_{\ell,p})$.

\paragraph{Frame-resolution rhythm and identity paths.}
Patch pooling can dilute short rhythm events. We therefore also add
\[
 r_{i,s}^{\mathrm{frame}}
   =\alpha_{i,s}^{\mathrm{rh}}W_re_i^{\mathrm{rh}}
\]
to the encoded motion frames before the trunk, bypassing condition pooling.
The projection $W_r$ starts at zero. A separate zero-initialized identity
projection is added to the diffusion-time token. For classifier-free guidance,
the unconditional branch replaces the base with a learned null embedding and
closes all residual gates; the identity-token path is retained. Guidance
therefore changes the speech-dependent update while keeping an explicit
speaker anchor.

\paragraph{Signed proposal--consensus injection.}
At each stage, let $h$ be a patch-level motion feature and $f$ its routed
condition. FiLM first modulates the motion feature:
\[
 \tilde{h}=h\odot(1+\tanh\gamma)+\tanh\beta.
\]
A bounded proposal $p$ is predicted from the modulated state and condition:
\[
 p=\tanh\!\left(P([\tilde{h};f])\right).
\]
The consensus network observes both the proposal and the same motion state:
\begin{equation}
\label{eq:dgn_consensus}
 g=\tanh\!\left(C([p;\tilde{h}])\right),\qquad
 h^{+}=\tilde{h}+\lambda_{\ell}(g\odot p).
\end{equation}
The consensus gate is signed: each component can preserve, suppress, or
reverse a proposed update. Its final linear layer is initialized to zero,
so proposal injection is initially inactive even though the learnable
stage scale $\lambda_{\ell}$ starts at one. The updated patches are then
processed by the temporal Transformer. This distinguishes modality selection,
which determines what speech information is available, from consensus
injection, which determines how that information changes the current motion.

\subsection{Matched-Noise Constraint Injection}
\begin{figure}
    \centering
    \includegraphics[width=\columnwidth]{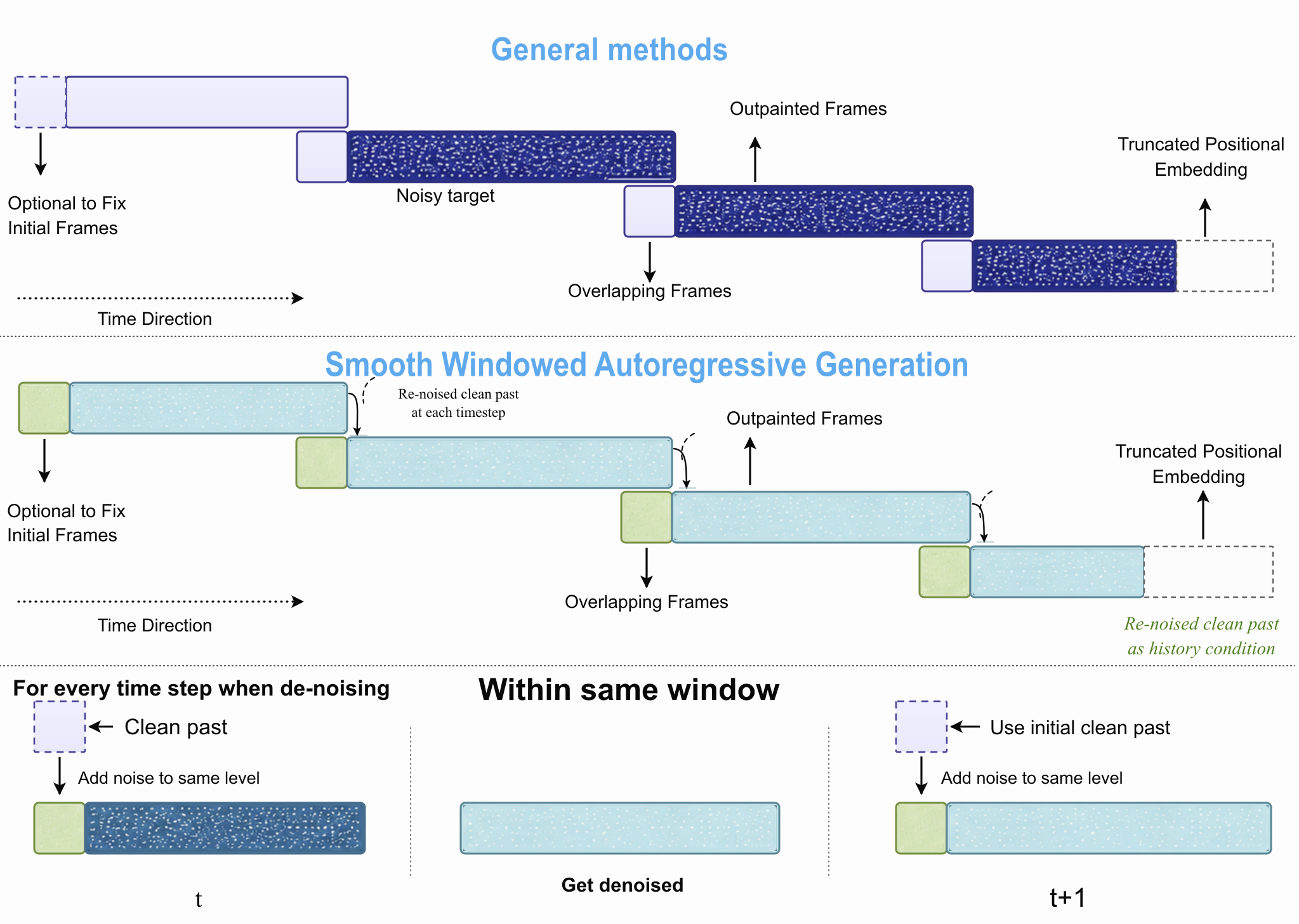}
    \caption{
    Matched-noise window continuation. Instead of stitching independently
    generated windows, DynaConTalk re-noises the committed clean history to the
    current diffusion level and injects it into the overlap region at every
    reverse step.
    }
    \Description{The window-continuation diagram showing clean history
    re-noised and injected into each reverse diffusion step.}
    \label{fig:windowed_sampling}
\end{figure}
We use a shared matched-noise constraint
mechanism~\cite{Lugmayr_2022_CVPR} for long-form window continuation, sparse
keypose editing, and reference-guided motion control. Constraints act directly
on the noisy wavelet sample, so the same denoiser and noise schedule can be
used across these tasks.

Let $x_0^c$ be a clean constraint in wavelet
space and $m$ be a soft mask over time, body part, and wavelet band.
Constraint coefficients use the same training-set normalization as the
diffusion target. At reverse
diffusion step $t$, we re-noise the constraint to the current diffusion level:
\begin{equation}
\label{eq:renoise}
    x_t^c=
    \sqrt{\bar{\alpha}_t}x_0^c+
    \sqrt{1-\bar{\alpha}_t}\epsilon ,
\end{equation}
and inject it into the current sample:
\begin{equation}
\label{eq:renoise_mask}
    \tilde{x}_t=(1-m)\odot x_t+m\odot x_t^c .
\end{equation}
The denoiser then predicts $\hat{x}_0$ from $\tilde{x}_t$, so constraints and
generated content are observed at a consistent noise level.

\paragraph{Seamless Window Transition.}
We now describe how we can apply this approach for seamlessly transitioning between motions synthesized per window.
Fig.~\ref{fig:windowed_sampling} top illustrates a standard windowed outpainting scheme~\cite{chen2024diffsheg}, during the diffusion process, the beginning of each new window is replaced with the ending frames of the previous window as conditioning, thereby enabling the generation of arbitrarily long motion sequences.

The middle row shows our approach. Instead of using the previous motion only
as an overlap between windows, we repeatedly re-noise the committed clean history as the constraint during the diffusion process.

The bottom row shows this operation in detail. At each reverse diffusion step,
we take the same clean history, add noise to match the current diffusion level as in Eq.~\ref{eq:renoise},
and inject it into the overlap region as in Eq.~\ref{eq:renoise_mask} by masking for the overlapping region. At the next step, we again start from
the original clean history, re-noise it to the new diffusion level, and repeat
the process. After sampling, only the newly generated future frames are
appended to the sequence. The released sampler uses 64-frame windows with
an eight-frame history and a 56-frame stride. It restores the overlap to the
clean history at the end of sampling. A four-frame dead blend further smooths
each join: the previous window's final velocity extrapolates the committed
motion, and a linear crossfade mixes that extrapolation with the new frames.
The history for the next window is taken from this committed sequence.

\paragraph{Keypose-guided editing.}
\begin{figure}
    \centering
    \includegraphics[width=\columnwidth]{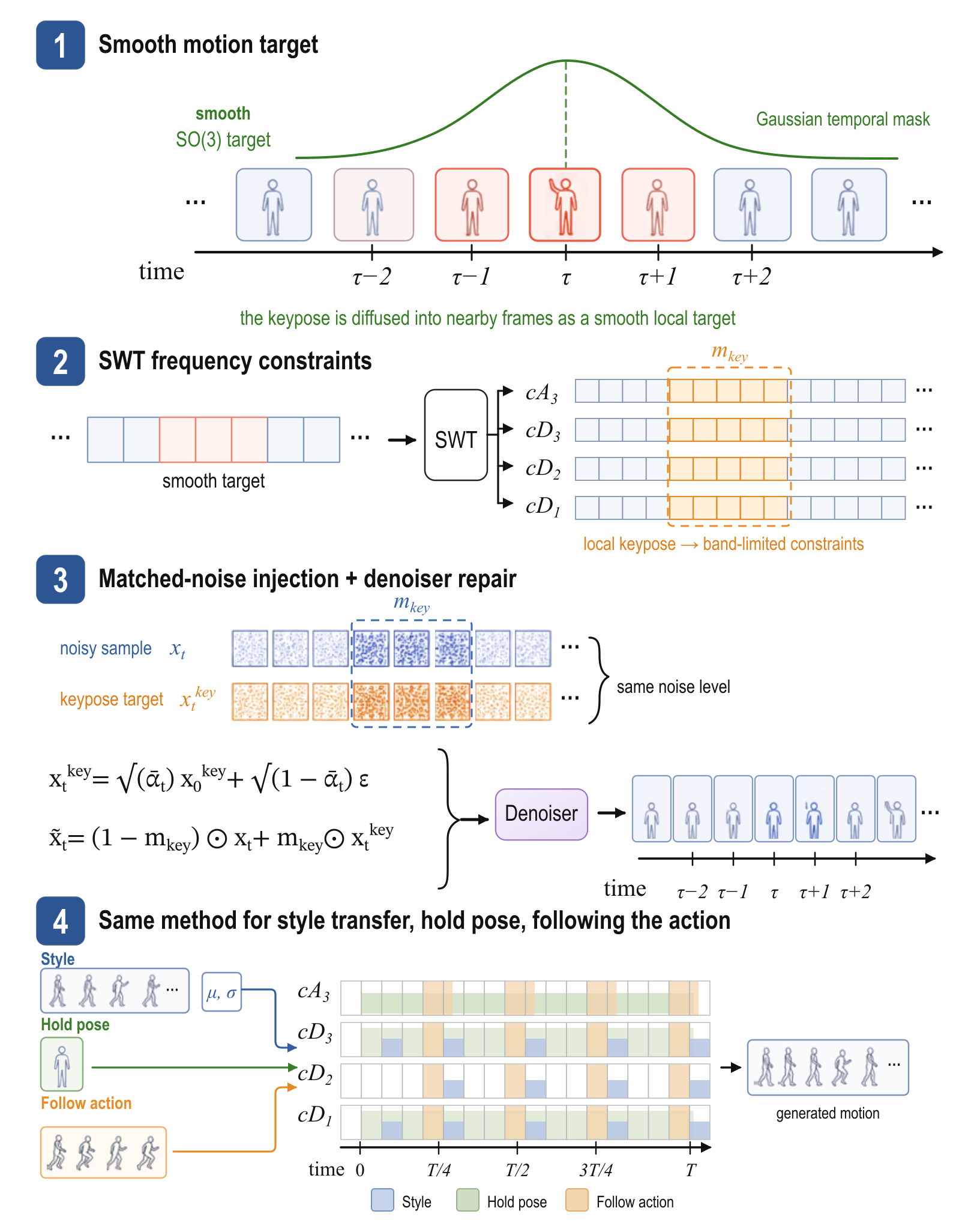}
    \caption{
    Keypose-guided wavelet repainting. A sparse user keypose is converted into a
    smooth SWT-domain target and injected through a time-part-band mask using
    the same matched-noise mechanism as history continuation.
    }
    \Description{The keypose-editing diagram showing a sparse pose target
    transformed to SWT space and injected for diffusion repair.}
    \label{fig:keypose_edit}
\end{figure}
As shown in Fig.~\ref{fig:keypose_edit}, for sparse pose editing, the user specifies
$\mathcal{K}=\{(\tau_i,J_i,y_i,w_i)\}$, where $\tau_i$ is the edited frame,
$J_i$ is the controlled body part, $y_i$ is the target pose, and
$w_i\in[0,1]$
is the
strength of the edit. Given the base generated motion $M^{\mathrm{base}}$, we construct a smooth
keypose target $M^{\mathrm{key}}$ around the edited frame. Let
$R^{\mathrm{base}}_{j,t}\in\mathrm{SO}(3)$ denote the rotation matrix of
joint $j$ at frame $t$ in $M^{\mathrm{base}}$,
$R^{\mathrm{tar}}_{j}\in\mathrm{SO}(3)$ the target rotation extracted from
the user-specified pose $y_i$, and $R^{\mathrm{key}}_{j,t}$ the resulting
smooth target rotation. For rotational keyposes, the target is formed by
interpolation in $\mathrm{SO}(3)$:
\[
R^{\mathrm{key}}_{j,t}
=
R^{\mathrm{base}}_{j,t}
\exp\!\left(
g_i(t)\log\!\left((R^{\mathrm{base}}_{j,t})^\top R^{\mathrm{tar}}_{j}\right)
\right), \quad j\in J_i ,
\]
where the temporal edit mask is defined as
\[
g_i(t)
=
w_i\exp\!\left(
-\frac{(t-\tau_i)^2}{2\sigma_i^2}
\right),
\]
where $\sigma_i$ controls the temporal width of the edit. Rotations outside the
controlled joint set $J_i$ remain unchanged. The complete sequence assembled from these target
rotations and represented in 6D rotation space is denoted by
$M^{\mathrm{key}}$. Let $\mathcal{W}$ denote the channel-wise temporal SWT
defined in Sec.~\ref{waveletdiff}. We transform $M^{\mathrm{key}}$ into SWT space and
construct a separable constraint mask
$m_{\mathrm{key}}$
over time, body parts, and wavelet
bands:
\[
    x^{\mathrm{key}}_0=\mathcal{W}(M^{\mathrm{key}}),
    \qquad
    m_{\mathrm{key}}=
    m_{\mathrm{time}}\otimes m_{\mathrm{part}}\otimes m_{\mathrm{band}} .
\]
In the SWT domain, the keypose constraint is no longer a frame-wise coordinate
replacement; instead, it is distributed over temporally aligned frequency bands,
anchoring low-frequency pose changes while allowing high-frequency gesture
details to be repaired by the denoising prior. Editing is then performed by
a second diffusion repair pass initialized from
the initial wavelet prediction. The user can control the edit strength, affected
body parts, and constrained wavelet bands. Since this uses the same
matched-noise injection as seamless window transition, the denoiser completes the
surrounding motion while following the sparse keypose constraint.
In the released editor, quaternion SLERP implements the rotation interpolation
above. Repainting starts halfway through the sampling schedule from a noised
copy of the base prediction. A fixed target-noise realization is reused across
the repaint steps, while history noise is resampled at each step. In overlap
frames, history takes priority over the edit mask. Windows with an empty edit
mask retain the base prediction. Edits are grouped into local clips, and the
repaired clips are crossfaded into the surrounding motion at their borders.

\paragraph{Reference-guided control.}
Through the wavelet representation, the same mechanism can be extended from
sparse keyposes to broader reference controls, including style control,
hold-pose guidance, and follow-action guidance.

For style control, let $R$ denote a reference motion, and let $p$ and $b$
index a body part and a wavelet band, respectively. We denote by
$\hat{x}_0^{p,b}$ and $x_R^{p,b}$ the corresponding part-band coefficients
of the current clean prediction and the reference motion,
$\mu_R^{p,b}$ and $\sigma_R^{p,b}$ the temporal mean and standard deviation of $x_R^{p,b}$, and $\mu_{\hat{x}}^{p,b}$ and
$\sigma_{\hat{x}}^{p,b}$ those of $\hat{x}_0^{p,b}$. We project the
current prediction toward the reference statistics by
\[
\begin{aligned}
\Pi_R(\hat{x}_0)^{p,b}
&=
\mu_R^{p,b}
+
\sigma_R^{p,b}
\frac{\hat{x}_0^{p,b}-\mu_{\hat{x}}^{p,b}}
{\sigma_{\hat{x}}^{p,b}+\epsilon}, \\
\hat{x}'_0
&=\hat{x}_0+\lambda m_R\odot
\left(\Pi_R(\hat{x}_0)-\hat{x}_0\right).
\end{aligned}
\]
where $\Pi_R$ denotes the reference-statistics projection, $\lambda$ controls its strength, and $m_R$ selects the affected time range, body parts, and wavelet bands.
For hold-pose and follow-action control, the selected pose is repeated or the reference action is looped/stretched into a clean target sequence $M^R=\mathcal{T}(R)$, transformed into SWT space, and injected as a time-part-band constraint:
\[
x_0^R=\mathcal{W}(M^R),\qquad
\tilde{x}_t
=
(1-m_R)\odot x_t
+
m_R\odot
\left(
\sqrt{\bar{\alpha}_t}x_0^R+
\sqrt{1-\bar{\alpha}_t}\epsilon
\right).
\]
Thus, style modifies part-band statistics, hold-pose anchors a long-range
postural target, and follow-action guides the motion along a reference action,
while the denoiser preserves speech compatibility and repairs unconstrained
details.

\subsection{Training Objective}

Both streams predict clean samples in normalized SWT coefficient space.
Using $\langle\cdot\rangle$ for an average over the indicated tensor entries,
the diffusion objective is
\[
 \mathcal{L}_{\mathrm{wav}}
 =\mathbb{E}_{x_0,s,\epsilon}
   \left[\left\langle(\hat{x}_0-x_0)^2\right\rangle\right].
\]
For the body stream, we de-normalize the coefficients, apply inverse SWT, and
compute SMPL-X joint positions. Position and velocity supervision are
\[
 \mathcal{L}_{\mathrm{geo}}
 =\left\langle(\mathrm{FK}(\hat{B})-\mathrm{FK}(B))^2\right\rangle,
 \qquad
 \mathcal{L}_{\mathrm{vel}}
 =\left\langle(\Delta\mathrm{FK}(\hat{B})-\Delta\mathrm{FK}(B))^2\right\rangle.
\]
Boundary supervision compares the first predicted future pose with the final
ground-truth history pose:
\[
 \mathcal{L}_{\mathrm{smo}}
 =\left\langle
  (\hat{B}^{\mathrm{fut}}_1-B^{\mathrm{hist}}_K)^2
 \right\rangle.
\]
The editable body recipe also applies a foot-contact loss. Predicted and
ground-truth local joints are placed in world coordinates under the root
trajectory used for training. Ground-truth foot displacement and height
identify the contact set $\mathcal{C}$, and the loss penalizes mean horizontal
foot displacement on those frames:
\[
 \mathcal{L}_{\mathrm{foot}}
 =\frac{1}{|\mathcal{C}|}
   \sum_{(i,j)\in\mathcal{C}}
   \|(\hat{p}_{i,j}-\hat{p}_{i-1,j})_{xz}\|_2.
\]
The contact thresholds are $0.01$ for per-frame horizontal displacement and
$0.05$\,m for height above the sequence's ground estimate; an empty contact
set contributes zero.

\paragraph{Sparse keypose supervision.}
For the editable body recipe, training clips have no keypose, one keypose,
or two to three keyposes with probabilities $0.5$, $0.25$, and $0.25$.
Keyposes select ground-truth motion at random future frames and body parts,
with Gaussian temporal masks of width $\sigma=6$ frames. The masked
ground-truth coefficients supply the training hint. The additional
supervision uses the soft mask as a weight:
\[
 \mathcal{L}_{\mathrm{kp}}
 =\frac{\sum_{i,d}m_{i,d}
       (\hat{x}_{0,i,d}-x_{0,i,d})^2}
       {\sum_{i,d}m_{i,d}+\epsilon}.
\]
The speech-only body and face recipes have no keypose supervision. At inference,
user keyposes replace these synthetic targets in the repainting mechanism.

\paragraph{Residual-gate regularization.}
The DGN residual gates are encouraged to open selectively through
\[
 \mathcal{L}_{\mathrm{gate}}
 =\mathbb{E}\!\left[
    \frac{1}{N|\mathcal{M}|}
    \sum_{i=1}^{N}\sum_{m\in\mathcal{M}}\alpha_{i,s}^m
   \right],\qquad
 \lambda_{\mathrm{gate}}=10^{-3}.
\]
Because the gates are nonnegative, this is a mean $L_1$ penalty. The full body
objective is
\[
\begin{aligned}
 \mathcal{L}_{B}
 &=\mathcal{L}_{\mathrm{wav}}
 +0.2\mathcal{L}_{\mathrm{geo}}
 +0.5\mathcal{L}_{\mathrm{vel}}
 +\mathcal{L}_{\mathrm{smo}}\\
 &\quad+\lambda_{\mathrm{foot}}\mathcal{L}_{\mathrm{foot}}
 +\lambda_{\mathrm{kp}}\mathcal{L}_{\mathrm{kp}}
 +10^{-3}\mathcal{L}_{\mathrm{gate}}.
\end{aligned}
\]
The editable recipe uses $\lambda_{\mathrm{foot}}=8$ and
$\lambda_{\mathrm{kp}}=1$; both are zero for the speech-only recipe.
The face objective is
$\mathcal{L}_{F}=\mathcal{L}_{\mathrm{wav}}+10^{-3}\mathcal{L}_{\mathrm{gate}}$.

\section{Experiments}
\label{sec:experiments}

We first evaluate the complete DynaConTalk system, then analyze the evidence
for each of its three core components: SWT motion representation, dynamic
proposal--consensus conditioning, and matched-noise constraint injection.

\subsection{Experimental Setup}
\label{sec:setup}

\paragraph{Datasets.}
We evaluate on two public holistic co-speech benchmarks,
BEAT2~\cite{liu2024emage} and SHOW~\cite{yi2023generating}. BEAT2 provides
motion-captured holistic body, hand, and facial motion, whereas SHOW contains
SMPL-X motion reconstructed from in-the-wild talking videos with synchronized
speech. We represent all motion at 30 fps.

\paragraph{Evaluation metrics.}
For all main comparisons, we use the official training, validation, and test splits. For body motion, Fr\'echet Gesture Distance (FGD)
~\cite{10.1145/3414685.3417838} measures the distribution of learned gesture
features, beat consistency (BC) measures speech--motion rhythm alignment, and
L1 diversity measures motion variation. Facial output is evaluated with vertex
MSE and face-vertex velocity distance (LVD), using the released EMAGE
evaluation tools. In Table~\ref{tab:beat2_quant}, the test-set ground truth
(GT) provides reference values for BC and diversity; generated values closer
to these references are preferred.

The speech-only body evaluation generates complete test sequences without
ground-truth motion history, keypose targets, or root-trajectory inputs.
FGD uses 6D joint rotations; BC and diversity use SMPL-X joint positions with
neutral body shape. BC excludes the first and last two seconds of each
sequence. For face evaluation, the first eight expression frames initialize
the history. SMPL-X vertices use the predicted expressions with the
ground-truth jaw pose and body shape. Thus, facial metrics assess expression
generation under this stated initialization and geometry protocol.

\paragraph{Implementation details.}
The body denoisers have four stages with 1024 channels and use
256-dimensional joint-band tokens, four spatial Transformer layers, two
cross-band layers, and temporal attention with 16 heads. The face denoiser
has four stages with 512 channels, uses temporal attention with eight heads,
and omits joint tokens. The condition embedding width is 256 for body and
128 for face. All recipes use three-level db6 SWT, a linear diffusion schedule
with 1000 training steps, 64-frame windows, and temporal patches of eight frames.

We optimize with AdamW, learning rate $2\times10^{-4}$, zero weight decay,
effective batch size 192, gradient clipping at $0.5$, and mixed bfloat16
precision. An exponential moving average with decay $0.995$ starts at
optimizer step 1000. The per-epoch cosine learning rate has minimum
$5\times10^{-6}$ and an initial period of 600 epochs. The period changes
to 160 at epoch 104 for the speech-only body recipe, 200 at epoch 136 for
the editable body recipe, and 158 at epoch 33 for the face recipe; it is
evaluated using the absolute epoch. Training ends at epochs 157, 176,
and 157, respectively (zero-based numbering). The reported speech-only
body checkpoint is epoch 157; the face checkpoint is epoch 150, selected
by validation expression MSE.

Classifier-free condition dropout has probability $0.2$, and partial
speech-modality dropout has probability $0.4$. Inference uses EMA weights,
10 second-order UniPC denoising steps, guidance scale 4.0, a 56-frame
window stride, and a four-frame dead blend. HuBERT features are extracted
in full precision and resampled from 50 to 30 fps; cached speech features
are stored in float16. The transcript is encoded with CLIP ViT-B/32.
Speaker identity and shape are supplied to both streams.

\subsection{Overall Holistic Generation Quality}
\begin{figure*}
  \centering
  \begin{minipage}[t]{0.38\textwidth}
    \centering
    \includegraphics[width=\linewidth]{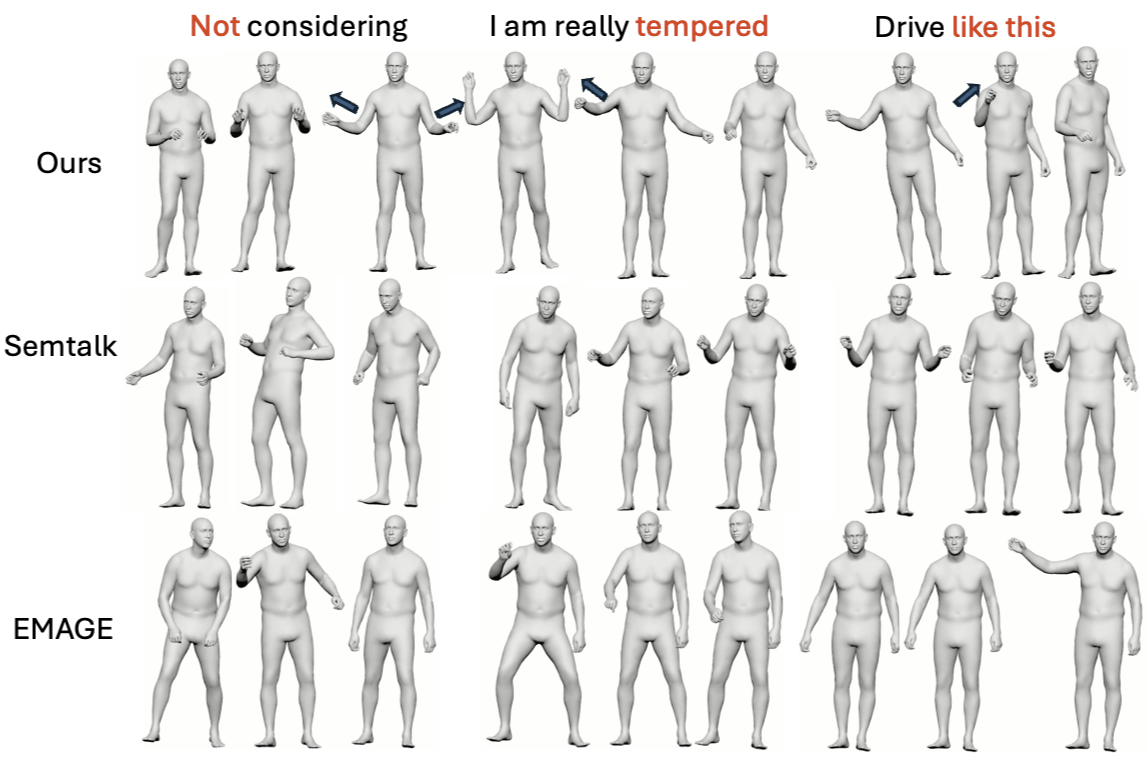}\par
    \small\textbf{(a) Body and gesture motion}
  \end{minipage}\hfill
  \begin{minipage}[t]{0.60\textwidth}
    \centering
    \includegraphics[width=\linewidth]{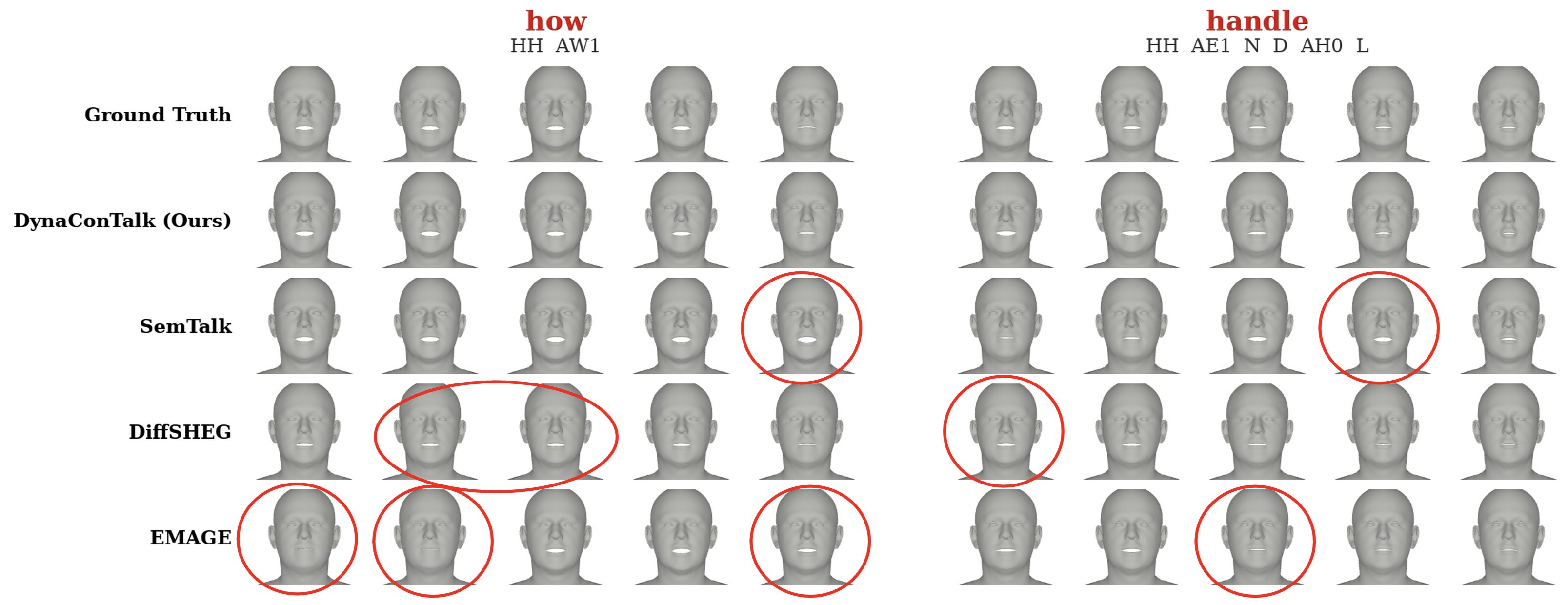}\par
    \small\textbf{(b) Facial expression}
  \end{minipage}
  \caption{Qualitative comparison on matched speech. The examples show an
  outward denial for ``not considering,'' raised hands for ``really tempered,''
  and a directional gesture for ``drive like this,'' together with coordinated
  body--hand motion and expressive facial detail.}
  \Description{Rendered motion frames compare DynaConTalk with EMAGE and SemTalk
  for body gesture and facial animation.}
  \label{fig:qualitative}
\end{figure*}

\paragraph{Quantitative and Qualitative Analysis}
As shown in Tables~\ref{tab:beat2_quant} and~\ref{tab:show_quant}, DynaConTalk achieves the best FGD
and facial reconstruction metrics among the compared methods on both BEAT2 and
SHOW. On BEAT2, the final DynaConTalk row in
Table~\ref{tab:beat2_quant} reports the speech-only body checkpoint at epoch
157 and the face checkpoint at epoch 150. Its raw metrics are FGD $0.175$,
BC $0.536$, diversity $7.85$, facial MSE $3.55\times10^{-8}$, and LVD
$1.32\times10^{-5}$. The speech-only row has lower FGD and LVD and is closer
to the test-set GT in both BC and diversity than the editable row, while the
editable row has lower facial MSE. These rows describe different generation
configurations. Figure~\ref{fig:qualitative} compares matched-speech body,
hand, and facial outputs with the strongest holistic baselines.

\label{sec:overall_quality}
\label{sec:sota}

\begin{table}[t]
  \centering
  \scriptsize
  \setlength{\tabcolsep}{3pt}
  \renewcommand{\arraystretch}{1.12}
  \caption{Quantitative comparison on BEAT2 All-Speakers Setting.
  We follow the evaluation protocol of RAG-GESTURE~\cite{mughal2025raggesture}.
  We report FGD$\times10^{-1}$,
  BC$\times10^{-1}$, MSE$\times10^{-8}$, and LVD$\times10^{-5}$.
  MSE and LVD evaluate facial motion. For BC and Div., closer to the test-set
  GT is better; lower is better for FGD, MSE, and LVD.
  The (edit) row uses the editable
  configuration; the final row reports the speech-only body model and the face
  model without keypose or trajectory controls. Best generated results are in bold.}
  \label{tab:beat2_quant}

  \resizebox{\columnwidth}{!}{
  \begin{tabular}{lccccc}
    \toprule
    Method
    & FGD$\downarrow$
    & BC
    & Div.
    & MSE$\downarrow$
    & LVD$\downarrow$ \\
    \midrule
    GT (test set)
      & -- & 4.77 & 7.29 & -- & -- \\
    \midrule
    CaMN~\cite{liu2022beat}
      & 5.12 & 2.00 & 5.58 & -- & -- \\
    Audio2Photoreal~\cite{ng2024audio}
      & 8.49 & 3.26 & 6.24 & -- & -- \\
    ReMoDiffuse~\cite{zhang2023remodiffuse}
      & 11.20 & 2.18 & 5.06 & -- & -- \\
    DSG~\cite{yang2023diffusestylegesture}
      & 11.74 & 7.34 & 11.12 & -- & -- \\
    HoloGest~\cite{cheng2025hologest}
      & 6.46 & 8.03 & 13.53 & -- & -- \\
    EMAGE~\cite{liu2024emage}
      & 6.92 & 2.84 & 6.06 & 6.908 & 1.453 \\
    RAG-GESTURE~\cite{mughal2025raggesture}
      & 4.87 & 5.14 & 9.94 & -- & -- \\
    SemTalk~\cite{zhang2025semtalk}
      & 3.56 & \textbf{5.10} & 8.409 & 4.439 & 1.435 \\
    DiffSHEG~\cite{chen2024diffsheg}
      & 9.43 & 2.25 & \textbf{7.23} & 7.665 & 8.673 \\
    \midrule
    DynaConTalk (edit)
      & 2.97
      & 6.83
      & 8.09
      & \textbf{3.413}
      & 1.422 \\
    DynaConTalk
      & \textbf{1.75}
      & 5.36
      & 7.85
      & 3.55
      & \textbf{1.32} \\
    \bottomrule
  \end{tabular}
  }

  \caption{Quantitative SHOW comparison. We report FGD$\times10^{-1}$, BC$\times10^{-1}$, MSE$\times10^{-8}$, and LVD$\times10^{-5}$. MSE and LVD evaluate facial motion. Best results are in bold.}
  \label{tab:show_quant}

  \resizebox{\columnwidth}{!}{
  \begin{tabular}{lccccc}
    \toprule
    Method
    & FGD$\downarrow$
    & BC$\uparrow$
    & Div.$\uparrow$
    & MSE$\downarrow$
    & LVD$\downarrow$ \\
    \midrule
    FaceFormer~\cite{fan2022faceformer}
      & -- & -- & -- & 138.1 & 43.69 \\
    CodeTalker~\cite{xing2023codetalker}
      & -- & -- & -- & 140.7 & 45.84 \\
    CaMN~\cite{liu2022beat}
      & 22.12 & 7.712 & 10.37 & -- & -- \\
    DSG~\cite{yang2023diffusestylegesture}
      & 24.84 & 8.027 & 10.23 & -- & -- \\
    LivelySpeaker~\cite{zhi2023livelyspeaker}
      & 32.17 & 7.844 & 10.14 & -- & -- \\
    Habibie et al.~\cite{habibie2021learning}
      & 27.22 & 8.209 & 8.541 & 145.6 & 47.35 \\
    TalkSHOW~\cite{yi2023generating}
      & 24.43 & 8.249 & 10.98 & 139.6 & 45.17 \\
    EMAGE~\cite{liu2024emage}
      & 22.12 & 8.280 & \textbf{12.46} & 136.1 & 42.44 \\
    SemTalk~\cite{zhang2025semtalk}
      & 20.18 & 8.304 & 11.36 & 134.1 & 39.15 \\
    DiffSHEG~\cite{chen2024diffsheg}
      & 24.87 & 8.061 & 10.79 & 139.0 & 45.77 \\
    \midrule
    DynaConTalk
      & \textbf{14.25}
      & \textbf{8.313}
      & 11.45
      & \textbf{105.41}
      & \textbf{28.59} \\
    \bottomrule
  \end{tabular}
  }

\end{table}

\begin{figure}
    \centering
    \includegraphics[width=\columnwidth]{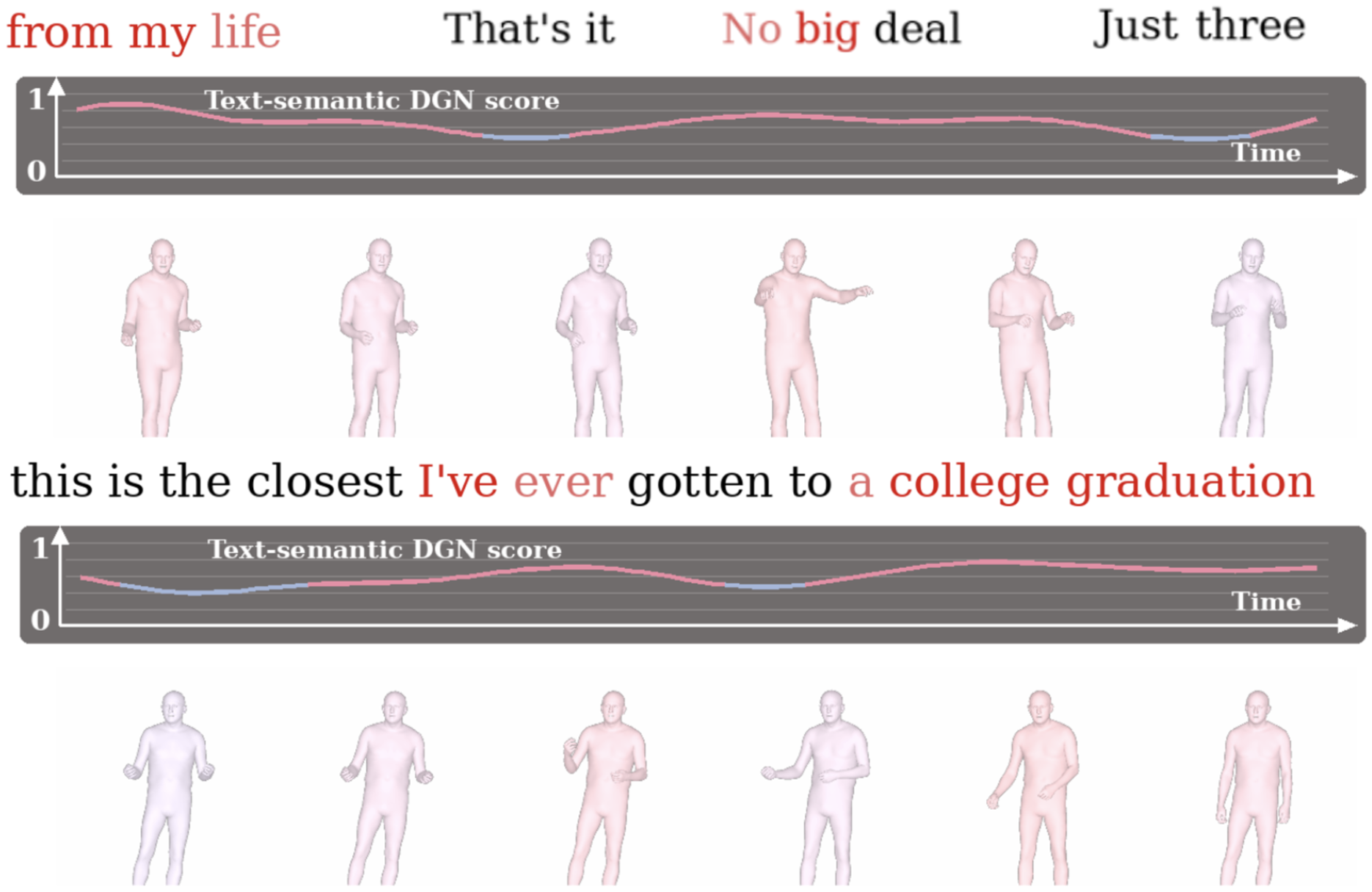}
    \caption{
Qualitative study on DGN. It can be observed that DGN increases the weight assigned to lexical information when semantically salient words occur.}
    \Description{Two speech examples show text-semantic DGN scores over time
    above corresponding sequences of generated body poses. Highlighted words
    coincide with increased lexical-conditioning weights and gesture changes.}
    \label{fig:Qualitative Analysis DGN}
\end{figure}
\begin{figure*}[t]
  \centering
  \includegraphics[width=\textwidth]{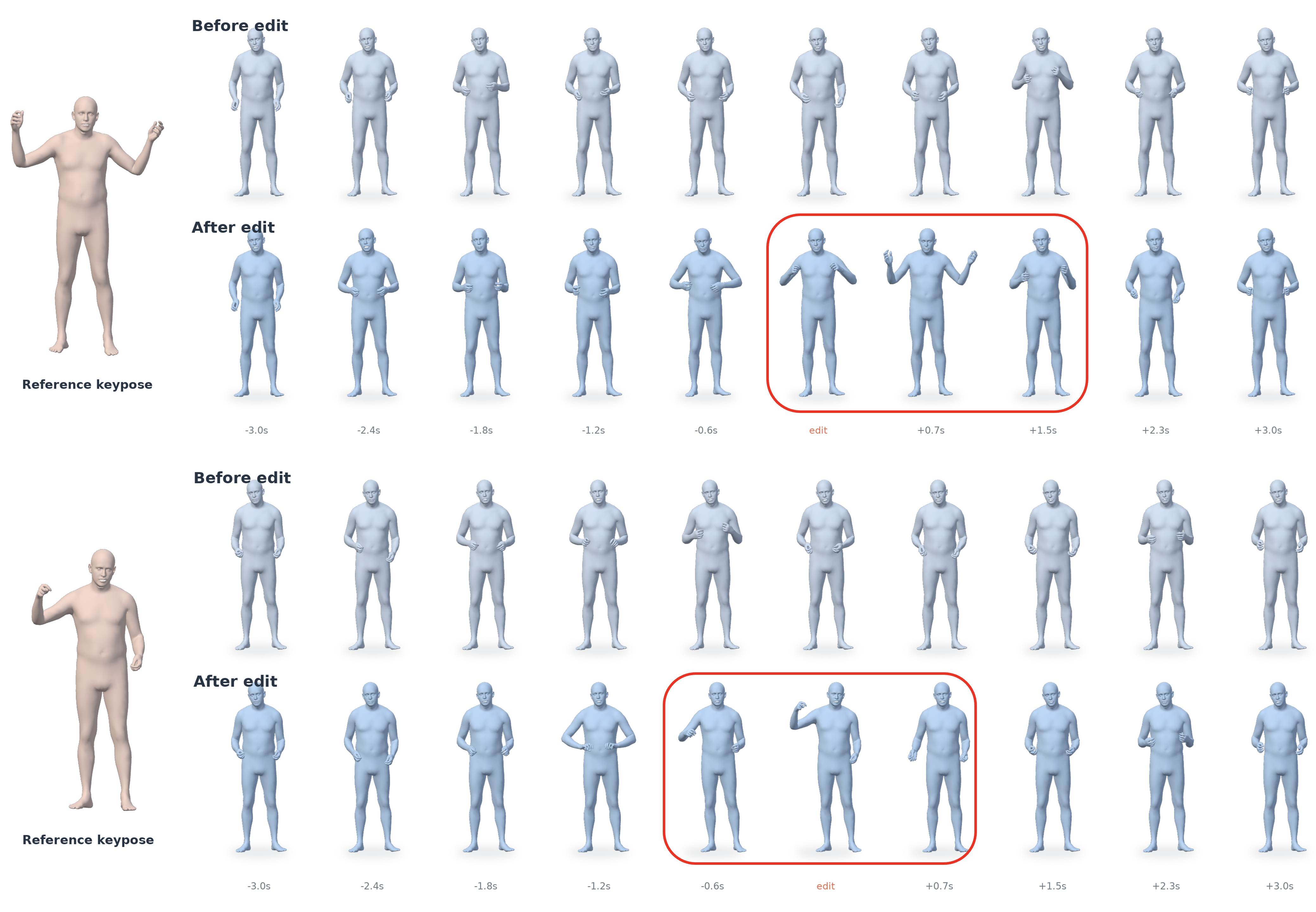}
  \caption{
  Reference-keypose motion editing. Given a target keypose and a specified edit
  time, DynaConTalk modifies the local motion to follow the reference while
  preserving the surrounding motion context. Red boxes highlight the regions
  most affected by the edit.
  }
  \Description{
  Two examples of reference-keypose motion editing. Each example shows a target
  reference keypose together with motion sequences before and after editing.
  Red boxes highlight the temporal regions modified around the specified edit
  time.
  }
  \label{fig:keypose_editing}
\end{figure*}

\paragraph{Reference-keypose editing.}
As shown in Fig.~\ref{fig:Qualitative Analysis DGN}, by directly learning the compatibility between motion and different conditioning cues, DGN can dynamically adjust gestures at appropriate moments. Figure~\ref{fig:keypose_editing} further demonstrates the controllability of
DynaConTalk through reference-keypose editing. Given a target pose and a
specified edit time, the model adjusts the local motion to follow the reference
while preserving the surrounding motion context. The examples show that the
edited gesture is introduced locally and transitions smoothly to the original
motion before and after the edited interval.

\paragraph{User study.}
Thirty participants rated 15 matched speech cases with anonymized outputs
from EMAGE, SemTalk, and DynaConTalk. Table~\ref{tab:user_study} reports that DynaConTalk
receives the highest mean score in realism, semantic congruence, synchrony, and
diversity. This study evaluates the complete systems---representation,
conditioning, and holistic output together.

\begin{table}[t]
\centering
\caption{User study on BEAT2. Mean ratings; higher is better.}
\label{tab:user_study}
\small
\setlength{\tabcolsep}{4.5pt}
\begin{tabular}{lrrrr}
\toprule
Method & Realism & Semantic & Synchrony & Diversity \\
\midrule
EMAGE & 1.67 & 1.89 & 1.76 & 1.83 \\
SemTalk & 2.05 & 1.95 & 2.05 & 2.05 \\
DynaConTalk & \textbf{2.30} & \textbf{2.15} & \textbf{2.21} & \textbf{2.15} \\
\bottomrule
\end{tabular}
\end{table}

\paragraph{Foot--ground stability.}
Under the CAMDM protocol~\cite{chen2024tamingdiffusionprobabilisticmodels},
DynaConTalk reduces normalized foot sliding to less than half that of the best
baseline and also reaches the lowest absolute and 95th-percentile sliding
(Table~\ref{tab:foot_sliding}).

\begin{table}[t]
\centering
\caption{Foot stability on the BEAT2 test set under the CAMDM protocol.}
\label{tab:foot_sliding}
\small
\setlength{\tabcolsep}{5pt}
\begin{tabular}{lrrrr}
\toprule
Method & Sliding$\downarrow$ & Abs.$\downarrow$ & P95$\downarrow$ & Contact \\
\midrule
EMAGE & 2.2040 & .01328 & .01908 & .0008 \\
SemTalk & 1.6942 & .00555 & .01507 & .0109 \\
DynaConTalk & \textbf{.7751} & \textbf{.00233} & \textbf{.00613} & .0933 \\
\bottomrule
\end{tabular}
\end{table}

\subsection{Full-System Ablation Analysis}
\label{sec:ablations}
We conducted ablation experiments on the single-person (speaker2) subset of the BEAT2 dataset, splitting the data into training, validation, and test sets at a ratio of 8:1:1.
\paragraph{Main component attribution.}
Table~\ref{tab:core_ablation} reports the core-component ablations under the
same full-system metrics.

\begin{table}[b]
\centering
\caption{Core-component attribution on BEAT2 speaker 2. $\Delta$FGD is relative to the full model.}
\label{tab:core_ablation}
\scriptsize
\setlength{\tabcolsep}{3.3pt}
\renewcommand{\arraystretch}{1.12}
\begin{tabularx}{\columnwidth}{@{}Xrrrr@{}}
\toprule
Variant & FGD$\downarrow$ & $\Delta$FGD & BC$\uparrow$ & L1-div.$\uparrow$ \\
\midrule
Full model & 3.107 & -- & 7.853 & 11.0900 \\
w/o DGN & 4.465 & +1.358 & 7.783 & 11.7689 \\
w/o consensus & 3.951 & +.844 & 7.911 & 11.6772 \\
DGN $\rightarrow$ cross-attn. & 3.747 & +.640 & 7.352 & 10.8291 \\
w/o SWT & 3.482 & +.375 & 7.717 & 12.1563 \\
\bottomrule
\end{tabularx}
\end{table}

Removing DGN causes the largest FGD degradation, followed by removing the
consensus operation, while replacing DGN with cross-attention also performs
worse than the full model (Table~\ref{tab:core_ablation}). Removing SWT has
a smaller effect on FGD; its contribution to fine-scale dynamics is examined
directly in Sec.~\ref{sec:optimization_geometry}.

\paragraph{Backbone and sampling choices.}
Temporal downsampling has the largest effect among the tested backbone
components, while the spatial Transformer,
band-specific heads, and proposed window sampling also improve generation
quality (Table~\ref{tab:backbone_ablation}).
\begin{table}[t]
\centering
\caption{Backbone and sampling ablations on BEAT2 speaker 2. $\Delta$FGD is relative to
the full model.}
\label{tab:backbone_ablation}
\scriptsize
\setlength{\tabcolsep}{3.3pt}
\begin{tabularx}{\columnwidth}{@{}Xrrrr@{}}
\toprule
Variant & FGD$\downarrow$ & $\Delta$FGD & BC$\uparrow$ & L1-div.$\uparrow$ \\
\midrule
w/o downsampling & 3.643 & +.536 & 7.572 & 11.6865 \\
w/o spatial Transformer & 3.445 & +.338 & 7.581 & 11.6986 \\
w/o band-specific heads & 3.411 & +.304 & 7.620 & 11.5354 \\
Smooth window sampling & 3.377 & +.270 & 7.573 & 11.9420 \\
Full model & 3.107 & -- & 7.853 & 11.0900 \\
\bottomrule
\end{tabularx}
\end{table}

\paragraph{Condition set.}
Table~\ref{tab:condition_ablation} evaluates the contribution of individual
conditioning inputs starting from the complete annotation set. Removing CLIP
emotion or word-index cues improves FGD by $0.108$ and $0.085$, respectively,
showing that including every available annotation is not necessarily beneficial.

Rhythm and content-related cues affect different aspects of generation.
Removing rhythm causes the largest drop in BC while leaving FGD nearly unchanged,
indicating that it primarily supports speech--motion synchronization. In contrast,
removing mel or HuBERT increases FGD by $0.047$ and $0.088$, respectively,
whereas removing CLIP text has only a marginal effect on FGD. These results
motivate selecting the conditioning set according to its measured contribution
rather than using all available inputs.

\begin{table}[t]
\centering
\caption{Condition-input ablations on BEAT2 speaker 2. $\Delta$FGD is relative to
All conditions; negative values indicate improvement.}
\label{tab:condition_ablation}
\scriptsize
\setlength{\tabcolsep}{3.3pt}
\begin{tabularx}{\columnwidth}{@{}Xrrrr@{}}
\toprule
Variant & FGD$\downarrow$ & $\Delta$FGD & BC$\uparrow$ & L1-div.$\uparrow$ \\
\midrule
All conditions & 3.325 & -- & 7.681 & 11.8578 \\
w/o CLIP emotion & 3.217 & $-.108$ & 7.721 & 11.3341 \\
w/o word index & 3.240 & $-.085$ & 7.729 & 11.3781 \\
w/o rhythm & 3.304 & $-.021$ & 7.428 & 11.3719 \\
w/o mel & 3.372 & +.047 & 7.766 & 10.7499 \\
w/o HuBERT semantics & 3.413 & +.088 & 7.724 & 11.5954 \\
w/o CLIP text & 3.305 & $-.020$ & 7.720 & 11.5920 \\
Full selected set & 3.107 & $-.218$ & 7.853 & 11.0900 \\
\bottomrule
\end{tabularx}
\end{table}

\paragraph{Wavelet family and level.}
Table~\ref{tab:wavelet_config_ablation} separates filter choice from
decomposition depth. Three-level db6 gives the strongest complete result, but
several other three-level families remain competitive.

\begin{table}[t]
\centering
\caption{Wavelet-family and decomposition-level ablations on BEAT2 speaker 2.
$\Delta$FGD is relative to the full model.}
\label{tab:wavelet_config_ablation}
\scriptsize
\setlength{\tabcolsep}{3.3pt}
\begin{tabularx}{\columnwidth}{@{}Xrrrr@{}}
\toprule
Variant & FGD$\downarrow$ & $\Delta$FGD & BC$\uparrow$ & L1-div.$\uparrow$ \\
\midrule
sym4 L3 & 3.162 & +.055 & 7.672 & 11.8648 \\
bior2.8 L3 & 3.145 & +.038 & 7.581 & 11.6986 \\
coif5 L3 & 3.348 & +.241 & 7.658 & 11.7614 \\
bior4.4 L3 & 3.199 & +.092 & 7.676 & 11.8984 \\
db6 L1 & 3.341 & +.234 & 7.698 & 11.9722 \\
db6 L2 & 3.178 & +.071 & 7.621 & 11.9938 \\
db6 L4 & 3.527 & +.420 & 7.262 & 11.4232 \\
w/o wavelet & 3.482 & +.375 & 7.717 & 12.1563 \\
Full: db6 L3 & 3.107 & -- & 7.853 & 11.0900 \\
\bottomrule
\end{tabularx}
\end{table}

The standard metrics select db6 L3 for the complete model. The targeted
multi-family study below then asks a different question: whether the dynamic
representation effect persists across filters even when their aggregate FGD
values are similar.

\subsection{Analysis of the SWT Motion Representation}
\subsubsection{Multiscale Structure of Natural Co-Speech Motion}
\label{sec:aligned_structure}

\begin{figure*}[t]
\centering
\includegraphics[width=\textwidth]{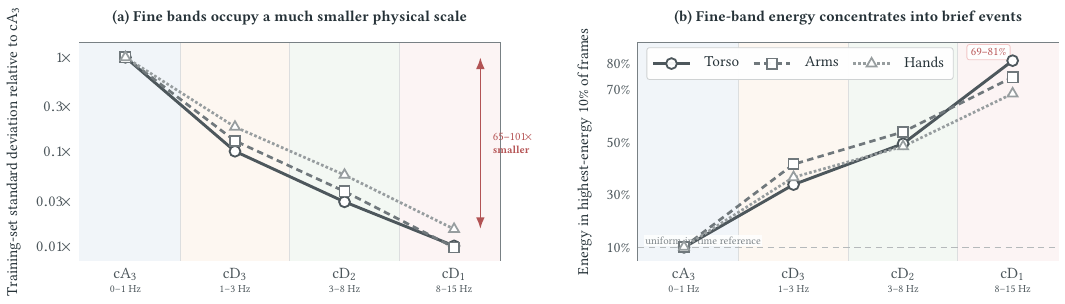}
\caption{
Multiscale structure of ground-truth co-speech motion on the BEAT2 training
set.
(a) Standard deviation of each SWT band relative to $\mathrm{cA}_3$.
(b) Fraction of band energy contained in the highest-energy 10\% of frames.
Across torso, arms, and hands, finer bands have progressively smaller physical
amplitudes while becoming more temporally concentrated: $\mathrm{cD}_1$ is
65--101$\times$ smaller in standard deviation than $\mathrm{cA}_3$, yet
69--81\% of its energy is concentrated in only 10\% of frames.
}
\Description{Three groups of plots compare the cA3, cD3, cD2, and cD1 wavelet
bands for torso, arms, and hands. Finer bands have progressively smaller
physical amplitudes and greater temporal energy concentration.}
\label{fig:band_structure}
\end{figure*}

We first examine the multiscale structure of ground-truth co-speech motion
on the BEAT2 training set. As shown in Fig.~\ref{fig:band_structure}, the four
SWT bands capture progressively faster temporal dynamics, ranging from slow
postural variation in $\mathrm{cA}_3$ to rapid local motion in
$\mathrm{cD}_1$.

More importantly, finer-scale motion differs substantially from coarse motion
in both amplitude and temporal distribution. Across the torso, arms, and
hands, the standard deviation of $\mathrm{cD}_1$ is 65--101$\times$ smaller
than that of $\mathrm{cA}_3$. At the same time, 69--81\% of its energy is
concentrated in the highest-energy 10\% of frames. Thus, fine-scale motion is
not only low in amplitude, but also highly localized in time. These brief
motion events can therefore contribute little to an aggregate raw-coordinate
loss despite being important to natural motion dynamics. This motivates the use of SWT to explicitly separate and model fine-scale
motion that would otherwise be underweighted in raw-coordinate learning.

\subsubsection{Representation-Space Error Analysis}
\label{sec:optimization_geometry}
The previous subsection established that SWT separates motion into distinct temporal scales. A natural question then arises: does this structured representation also make the generation process more robust to prediction errors? To answer this, we conduct a controlled perturbation experiment that isolates the effect of the representation space itself, independent of the denoiser or conditioning.

We take ground-truth motion sequences from the BEAT2 test set and add a small amount of Gaussian noise to either:
    the Raw6D coordinate space (direct perturbation of 6D rotations), or
    the SWT coefficient space (perturbation applied to each wavelet band, followed by inverse SWT to reconstruct the motion).
The noise magnitude is calibrated so that the resulting motion exhibits a similar overall root-mean-square error (RMSE) in pose space across both conditions. This ensures a fair comparison: we are not asking which representation preserves the pose better, but which representation produces more natural motion dynamics from the same level of pose error.

Let $E=\widetilde{M}-M$ denote the difference between the perturbed and original motion. Since Raw6D and SWT are matched to have the same pose RMSE, we focus on how this error changes over time. We measure temporal distortion using the second-order temporal difference of $E$, where a larger value indicates stronger frame-to-frame jitter.

Under this matched-error setting, Raw6D yields a temporal residual of $44.10$, while SWT yields only $4.67$, corresponding to an $89.4\%$ reduction. This shows that, even with the same overall pose error, perturbations in SWT space lead to much smoother and less disruptive motion dynamics than perturbations in Raw6D space.

\paragraph{Dynamic-distribution validation.}
\label{sec:fulltest}

\begin{figure}
    \centering
    \includegraphics[width=\columnwidth]
    {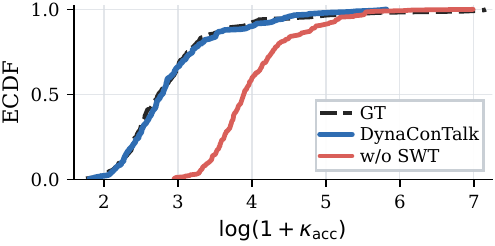}
\caption{
    Full-test distributions of $\log(1+\kappa_{\mathrm{acc}})$.
    DynaConTalk closely follows the GT ECDF, whereas Raw6D is strongly shifted
    toward higher acceleration kurtosis, indicating more impulsive and
    heavy-tailed dynamics.
}
    \Description{
The DynaConTalk curve closely follows the GT curve.
    }
    \label{fig:fulltest}
\end{figure}

To evaluate whether SWT better preserves the brief, high-acceleration events
characteristic of natural motion, we compare the distribution of upper-body
joint-acceleration kurtosis, denoted by $\kappa_{\mathrm{acc}}$. We report its log-transformed value
$\log(1+\kappa_{\mathrm{acc}})$ and compare generated and GT distributions
using the 1-Wasserstein distance ($W_1$) and Kolmogorov--Smirnov (KS)
statistics. Under the same DGN conditioning, DynaConTalk closely matches the GT
distribution, with $W_1=0.0887$, compared with $1.1217$ for
Raw6D---a 92.1\% reduction
(Table~\ref{tab:fulltest}, Fig.~\ref{fig:fulltest}). A two-sample KS test also
does not reject distributional agreement with GT ($p=0.57$). Because this
evaluation uses joint positions after kinematic conversion, the representation
advantage also carries over to physical motion.

\begin{table}
\centering
\caption{Full-test upper-body joint-acceleration distribution.}
\label{tab:fulltest}
\small
\setlength{\tabcolsep}{4pt}
\begin{tabular}{lrrr}
\toprule
Method & Median & $W_1$ to GT $\downarrow$ & KS to GT $\downarrow$ \\
\midrule
GT & 2.7386 & 0 & 0 \\
DynaConTalk & 2.7901 & \textbf{0.0887} & \textbf{0.0679} \\
Raw6D & 3.8911 & 1.1217 & 0.7849 \\
EMAGE & 2.2619 & 0.5540 & 0.5472 \\
SemTalk & 2.1035 & 0.8015 & 0.6906 \\
\bottomrule
\end{tabular}
\end{table}

\paragraph{Why FGD Underrepresents Fine-Scale Improvements}
SWT substantially improves fine-scale dynamics, yet its removal changes FGD
only modestly. We therefore examine whether FGD is less sensitive to
high-frequency motion. First, we apply perturbations of the same magnitude at
different temporal frequencies and find that FGD responds about 20$\times$
less to 4--12~Hz perturbations than to 0.25--2~Hz perturbations. Second, in a
band-replacement test, replacing the predicted 3--15~Hz bands with GT leaves
99.51\% of FGD unchanged, whereas replacing only the 0--1~Hz band reduces the
remaining FGD to 9.14\%. These results show that FGD is mainly determined by
coarse, low-frequency motion and is relatively insensitive to the fine-scale
dynamics improved by SWT.

\subsection{Evaluation of Long-Form and Controllable Generation}
\label{sec:control_results}

\begin{figure}
    \centering
    \includegraphics[width=\columnwidth]{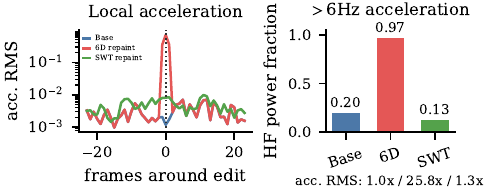}
    \caption{
Local smoothness around keypose editing.
    }
    \Description{A line plot compares local acceleration around an edited
    keypose for the base motion, Raw6D repainting, and SWT repainting. A bar
    chart shows that SWT repainting has a much smaller fraction of acceleration
    power above 6 Hz than Raw6D repainting.}
    \label{fig:keypose_acceleration_appendix}
\end{figure}

Table~\ref{tab:control_results} evaluates matched-noise constraint injection
for long-form continuation, keypose editing, and reference-guided control.
For window continuation, matched-noise history injection reduces boundary
position error by about one third and roughly halves velocity and acceleration
jumps relative to overlap replacement, improving continuity across generation
windows without a separate transition network. For keypose editing, SWT repaint
uses the same editing protocol as Raw6D repaint but reduces edit spill by
42\% while improving boundary continuity, indicating better localization of
the edited motion as shown in Fig.~\ref{fig:keypose_acceleration_appendix}. The same constraint mechanism supports reference-guided style, pose-hold, and follow-action control without retraining, reducing style distance by about one
third and nearly halving both pose-hold and motion-envelope errors relative to
uncontrolled generation.

\begin{figure}
    \centering
    \includegraphics[width=\columnwidth]{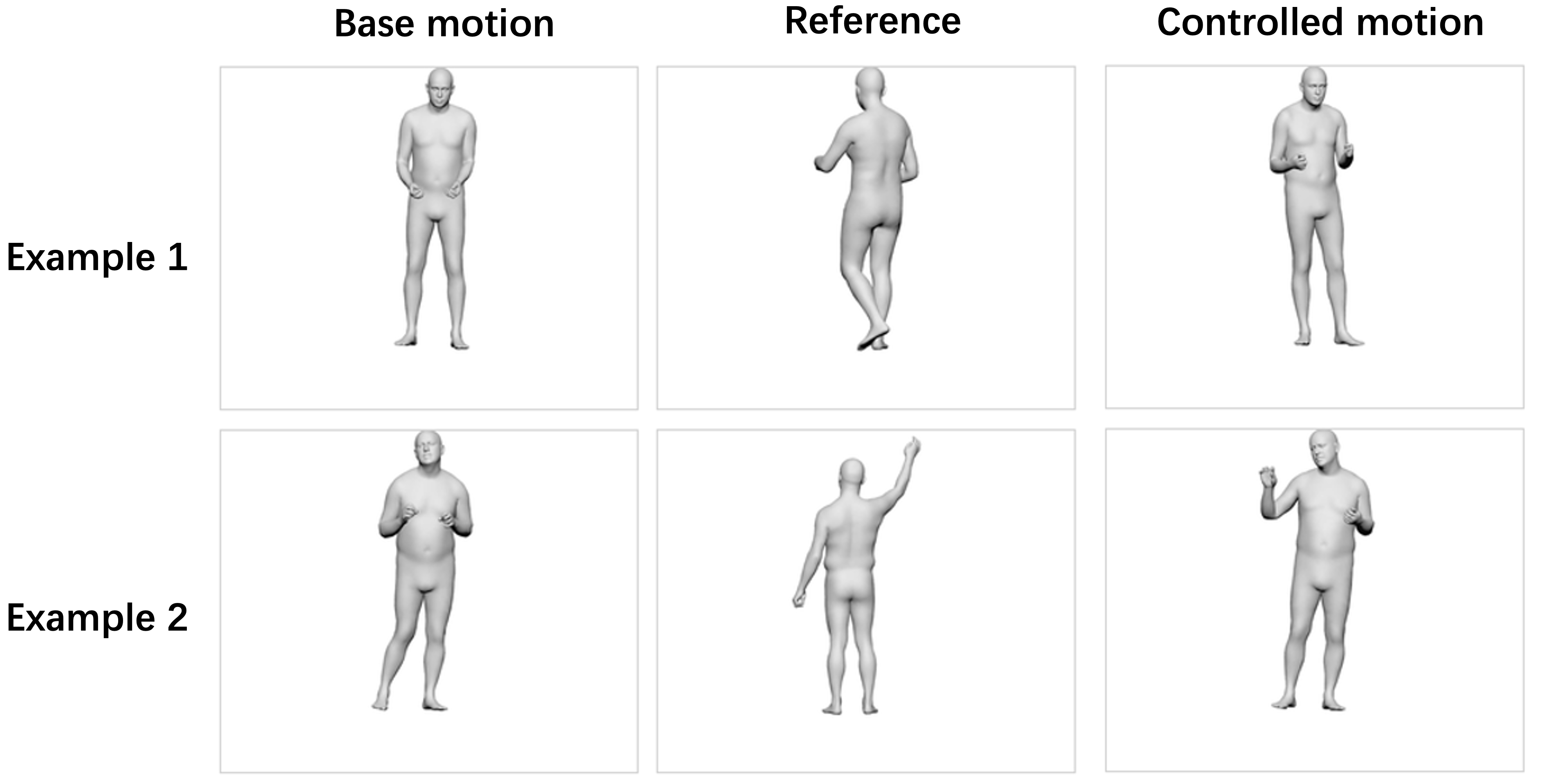}
    \caption{
Qualitative examples of reference-guided control.
    }
    \Description{Two examples compare a base motion, a reference pose, and the
    resulting controlled motion. The controlled characters adapt their body
    and arm gestures toward the references while retaining the base motions.}
    \label{fig:reference_control_qual}
\end{figure}

\begin{table}
\centering
\caption{Long-form and control evaluation. The comparison column is overlap
replacement for continuation, Raw6D repaint / DiffSHEG for keypose repair
(in that order), and the uncontrolled output for reference control. Raw6D
repaint uses our matched-noise protocol without SWT. Lower is better.}
\label{tab:control_results}
\footnotesize
\setlength{\tabcolsep}{3.3pt}
\begin{tabular}{@{}llrr@{}}
\toprule
Setting & Metric & Comparator(s) & Ours / ctrl. \\
\midrule
\multirow{3}{*}{Window}
 & Position & .0240 & \textbf{.0158} \\
 & Velocity & .0069 & \textbf{.0029} \\
 & Acceleration & .0034 & \textbf{.0017} \\
\midrule
\multirow{3}{*}{Keypose}
 & Edge acceleration & 6.35 / 6.46 & \textbf{5.89} \\
 & Edge velocity & 1.84 / 1.84 & \textbf{1.72} \\
 & Spill & 2.131 / 1.249 & \textbf{1.242} \\
\midrule
\multirow{3}{*}{Reference}
 & Style distance & .822 & \textbf{.537} \\
 & Hold RMS & .298 & \textbf{.148} \\
 & Follow error & .656 & \textbf{.360} \\
\bottomrule
\end{tabular}
\end{table}

Additional style and follow-action results are shown in
Fig.~\ref{fig:reference_control_qual} and the supplementary video.

\section{Limitations}
\label{sec:limitations}
Most holistic co-speech datasets provide limited
global locomotion, making root-trajectory generation weakly constrained by
speech alone. Future work will investigate larger-scale training and stronger
trajectory priors for more expressive whole-body motion. Generalization to more
diverse speakers and locomotion-heavy sequences also remains to be evaluated.

\section{Conclusion}

DynaConTalk addresses the loss of multiscale motion detail, static fusion of
heterogeneous speech cues, and discontinuities introduced by windowed
generation. It combines aligned SWT-domain diffusion, dynamic
state-aware residual gating with attention pooling and depth routing,
signed proposal--consensus conditioning, and matched-noise constraint
injection for holistic speech-driven motion. Experiments support improved
generation quality, while the controlled analyses associate dynamic shape with
the SWT representation, activity calibration with DGN, and smoother window
transitions and localized control with matched-noise constraints.

\begin{acks}
The authors thank all participants in the user study for their time and
valuable feedback. The authors also acknowledge the Interactive Content
Design Lab, Research Institute of Electrical Communication, Tohoku University,
for providing the facilities and computational resources that supported this
work.
This work was partially funded by the Research Grants Council of
Hong Kong (Ref: 17216826) and by the Innovation and Technology
Commission of the HKSAR Government under the ITSP-Platform
grant (Ref: ITS/335/23FP,  ITS/469/24FP).
\end{acks}

\bibliographystyle{dynacontalk-author-year}
\bibliography{sample-base}

@article{Pang_2023,
   title={Bodyformer: Semantics-guided 3D Body Gesture Synthesis with Transformer},
   volume={42},
   ISSN={1557-7368},
   url={http://dx.doi.org/10.1145/3592456},
   DOI={10.1145/3592456},
   number={4},
   journal={ACM Transactions on Graphics},
   publisher={Association for Computing Machinery (ACM)},
   author={Pang, Kunkun and Qin, Dafei and Fan, Yingruo and Habekost, Julian and Shiratori, Takaaki and Yamagishi, Junichi and Komura, Taku},
   year={2023},
   month={July}, pages={1--12} }

@inproceedings{zhang2025semtalk,
  title={Semtalk: Holistic co-speech motion generation with frame-level semantic emphasis},
  author={Zhang, Xiangyue and Li, Jianfang and Zhang, Jiaxu and Dang, Ziqiang and Ren, Jianqiang and Bo, Liefeng and Tu, Zhigang},
  booktitle={Proceedings of the IEEE/CVF International Conference on Computer Vision},
  pages={13761--13771},
  publisher={IEEE Computer Society},
  address={Los Alamitos, CA, USA},
  year={2025}
}

@inproceedings{fan2022faceformer,
  title={Faceformer: Speech-driven 3d facial animation with transformers},
  author={Fan, Yingruo and Lin, Zhaojiang and Saito, Jun and Wang, Wenping and Komura, Taku},
  booktitle={Proceedings of the IEEE/CVF Conference on Computer Vision and Pattern Recognition},
  pages={18770--18780},
  publisher={IEEE Computer Society},
  address={Los Alamitos, CA, USA},
  year={2022}
}

@inproceedings{xing2023codetalker,
  title={Codetalker: Speech-driven 3d facial animation with discrete motion prior},
  author={Xing, Jinbo and Xia, Menghan and Zhang, Yuechen and Cun, Xiaodong and Wang, Jue and Wong, Tien-Tsin},
  booktitle={Proceedings of the IEEE/CVF Conference on Computer Vision and Pattern Recognition},
  pages={12780--12790},
  publisher={IEEE Computer Society},
  address={Los Alamitos, CA, USA},
  year={2023}
}

@inproceedings{liu2022beat,
  title={Beat: A large-scale semantic and emotional multi-modal dataset for conversational gestures synthesis},
  author={Liu, Haiyang and Zhu, Zihao and Iwamoto, Naoya and Peng, Yichen and Li, Zhengqing and Zhou, You and Bozkurt, Elif and Zheng, Bo},
  booktitle={European conference on computer vision},
  pages={612--630},
  year={2022},
  publisher={Springer},
  address={Cham, Switzerland}
}

@inproceedings{yang2023diffusestylegesture,
  title={{DiffuseStyleGesture}: Stylized Audio-Driven Co-Speech Gesture Generation with Diffusion Models},
  author={Yang, Sicheng and Wu, Zhiyong and Li, Minglei and Zhang, Zhensong and Hao, Lei and Bao, Weihong and Cheng, Ming and Xiao, Long},
  booktitle={Proceedings of the Thirty-Second International Joint Conference on Artificial Intelligence, {IJCAI-23}},
  pages={5860--5868},
  year={2023},
  publisher={International Joint Conferences on Artificial Intelligence Organization},
  address={California, USA},
  doi={10.24963/ijcai.2023/650}
}

@inproceedings{ahuja2020style,
  title={Style transfer for co-speech gesture animation: A multi-speaker conditional-mixture approach},
  author={Ahuja, Chaitanya and Lee, Dong Won and Nakano, Yukiko I and Morency, Louis-Philippe},
  booktitle={Computer Vision--ECCV 2020: 16th European Conference, Glasgow, UK, August 23--28, 2020, Proceedings, Part XVIII 16},
  pages={248--265},
  year={2020},
  publisher={Springer},
  address={Cham, Switzerland}
}

@inproceedings{yi2023generating,
  title={Generating holistic 3d human motion from speech},
  author={Yi, Hongwei and Liang, Hualin and Liu, Yifei and Cao, Qiong and Wen, Yandong and Bolkart, Timo and Tao, Dacheng and Black, Michael J},
  booktitle={Proceedings of the IEEE/CVF Conference on Computer Vision and Pattern Recognition},
  pages={469--480},
  publisher={IEEE Computer Society},
  address={Los Alamitos, CA, USA},
  year={2023}
}

@inproceedings{liu2024emage,
  title={EMAGE: Towards Unified Holistic Co-Speech Gesture Generation via Expressive Masked Audio Gesture Modeling},
  author={Liu, Haiyang and Zhu, Zihao and Becherini, Giorgio and Peng, Yichen and Su, Mingyang and Zhou, You and Zhe, Xuefei and Iwamoto, Naoya and Zheng, Bo and Black, Michael J},
  booktitle={Proceedings of the IEEE/CVF Conference on Computer Vision and Pattern Recognition},
  pages={1144--1154},
  publisher={IEEE Computer Society},
  address={Los Alamitos, CA, USA},
  year={2024}
}

@inproceedings{chen2024diffsheg,
  title={Diffsheg: A diffusion-based approach for real-time speech-driven holistic 3d expression and gesture generation},
  author={Chen, Junming and Liu, Yunfei and Wang, Jianan and Zeng, Ailing and Li, Yu and Chen, Qifeng},
  booktitle={Proceedings of the IEEE/CVF Conference on Computer Vision and Pattern Recognition},
  pages={7352--7361},
  publisher={IEEE Computer Society},
  address={Los Alamitos, CA, USA},
  year={2024}
}

@book{kipp2005gesture,
  title={Gesture Generation by Imitation: From Human Behavior to Computer Character Animation},
  author={Kipp, Michael},
  year={2005},
  publisher={Dissertation.com / Universal Publishers},
  address={Boca Raton, FL, USA},
  isbn={1-58112-255-1}
}

@inproceedings{kopp2006towards,
  title={Towards a common framework for multimodal generation: The behavior markup language},
  author={Kopp, Stefan and Krenn, Brigitte and Marsella, Stacy and Marshall, Andrew N and Pelachaud, Catherine and Pirker, Hannes and Th{\'o}risson, Kristinn R and Vilhj{\'a}lmsson, Hannes},
  booktitle={Intelligent Virtual Agents: 6th International Conference, IVA 2006, Marina Del Rey, CA, USA, August 21-23, 2006. Proceedings 6},
  pages={205--217},
  year={2006},
  publisher={Springer},
  address={Berlin, Heidelberg}
}

@article{prendinger2004mpml,
  title={MPML: A markup language for controlling the behavior of life-like characters},
  author={Prendinger, Helmut and Descamps, Sylvain and Ishizuka, Mitsuru},
  journal={Journal of Visual Languages \& Computing},
  volume={15},
  number={2},
  pages={183--203},
  year={2004},
  publisher={Elsevier}
}

@inproceedings{cassell2001beat,
  title={Beat: the behavior expression animation toolkit},
  author={Cassell, Justine and Vilhj{\'a}lmsson, Hannes H{\"o}gni and Bickmore, Timothy W.},
  booktitle={Proceedings of the 28th annual conference on Computer graphics and interactive techniques},
  pages={477--486},
  publisher={Association for Computing Machinery},
  address={New York, NY, USA},
  year={2001}
}

@inproceedings{yoon2019robots,
  title={Robots learn social skills: End-to-end learning of co-speech gesture generation for humanoid robots},
  author={Yoon, Youngwoo and Ko, Woo-Ri and Jang, Minsu and Lee, Jaeyeon and Kim, Jaehong and Lee, Geehyuk},
  booktitle={2019 International Conference on Robotics and Automation (ICRA)},
  pages={4303--4309},
  year={2019},
  publisher={IEEE},
  address={Piscataway, NJ, USA}
}

@inproceedings{ye2022audio,
  title={Audio-driven stylized gesture generation with flow-based model},
  author={Ye, Sheng and Wen, Yu-Hui and Sun, Yanan and He, Ying and Zhang, Ziyang and Wang, Yaoyuan and He, Weihua and Liu, Yong-Jin},
  booktitle={European Conference on Computer Vision},
  pages={712--728},
  year={2022},
  publisher={Springer},
  address={Cham, Switzerland}
}

@inproceedings{habibie2021learning,
  title={Learning speech-driven 3d conversational gestures from video},
  author={Habibie, Ikhsanul and Xu, Weipeng and Mehta, Dushyant and Liu, Lingjie and Seidel, Hans-Peter and Pons-Moll, Gerard and Elgharib, Mohamed and Theobalt, Christian},
  booktitle={Proceedings of the 21st ACM International Conference on Intelligent Virtual Agents},
  pages={101--108},
  publisher={Association for Computing Machinery},
  address={New York, NY, USA},
  year={2021}
}

@article{alexanderson2023listen,
  title={Listen, denoise, action! audio-driven motion synthesis with diffusion models},
  author={Alexanderson, Simon and Nagy, Rajmund and Beskow, Jonas and Henter, Gustav Eje},
  journal={ACM Transactions on Graphics (TOG)},
  volume={42},
  number={4},
  pages={1--20},
  year={2023},
  publisher={ACM New York, NY, USA}
}

@inproceedings{zhu2023taming,
  title={Taming diffusion models for audio-driven co-speech gesture generation},
  author={Zhu, Lingting and Liu, Xian and Liu, Xuanyu and Qian, Rui and Liu, Ziwei and Yu, Lequan},
  booktitle={Proceedings of the IEEE/CVF Conference on Computer Vision and Pattern Recognition},
  pages={10544--10553},
  publisher={IEEE Computer Society},
  address={Los Alamitos, CA, USA},
  year={2023}
}

@inproceedings{bhattacharya2021speech2affectivegestures,
  title={Speech2affectivegestures: Synthesizing co-speech gestures with generative adversarial affective expression learning},
  author={Bhattacharya, Uttaran and Childs, Elizabeth and Rewkowski, Nicholas and Manocha, Dinesh},
  booktitle={Proceedings of the 29th ACM International Conference on Multimedia},
  pages={2027--2036},
  publisher={Association for Computing Machinery},
  address={New York, NY, USA},
  year={2021}
}

@inproceedings{liu2022learning,
  title={Learning hierarchical cross-modal association for co-speech gesture generation},
  author={Liu, Xian and Wu, Qianyi and Zhou, Hang and Xu, Yinghao and Qian, Rui and Lin, Xinyi and Zhou, Xiaowei and Wu, Wayne and Dai, Bo and Zhou, Bolei},
  booktitle={Proceedings of the IEEE/CVF Conference on Computer Vision and Pattern Recognition},
  pages={10462--10472},
  publisher={IEEE Computer Society},
  address={Los Alamitos, CA, USA},
  year={2022}
}

@inproceedings{liang2022seeg,
  title={Seeg: Semantic energized co-speech gesture generation},
  author={Liang, Yuanzhi and Feng, Qianyu and Zhu, Linchao and Hu, Li and Pan, Pan and Yang, Yi},
  booktitle={Proceedings of the IEEE/CVF Conference on Computer Vision and Pattern Recognition},
  pages={10473--10482},
  publisher={IEEE Computer Society},
  address={Los Alamitos, CA, USA},
  year={2022}
}

@inproceedings{zhi2023livelyspeaker,
  title={Livelyspeaker: Towards semantic-aware co-speech gesture generation},
  author={Zhi, Yihao and Cun, Xiaodong and Chen, Xuelin and Shen, Xi and Guo, Wen and Huang, Shaoli and Gao, Shenghua},
  booktitle={Proceedings of the IEEE/CVF International Conference on Computer Vision},
  pages={20807--20817},
  publisher={IEEE Computer Society},
  address={Los Alamitos, CA, USA},
  year={2023}
}

@inproceedings{ng2024audio,
  title={From audio to photoreal embodiment: Synthesizing humans in conversations},
  author={Ng, Evonne and Romero, Javier and Bagautdinov, Timur and Bai, Shaojie and Darrell, Trevor and Kanazawa, Angjoo and Richard, Alexander},
  booktitle={Proceedings of the IEEE/CVF Conference on Computer Vision and Pattern Recognition},
  pages={1001--1010},
  publisher={IEEE Computer Society},
  address={Los Alamitos, CA, USA},
  year={2024}
}

@inproceedings{liu2024towards,
  title={Towards Variable and Coordinated Holistic Co-Speech Motion Generation},
  author={Liu, Yifei and Cao, Qiong and Wen, Yandong and Jiang, Huaiguang and Ding, Changxing},
  booktitle={Proceedings of the IEEE/CVF Conference on Computer Vision and Pattern Recognition},
  pages={1566--1576},
  publisher={IEEE Computer Society},
  address={Los Alamitos, CA, USA},
  year={2024}
}

@article{ao2023gesturediffuclip,
  title={Gesturediffuclip: Gesture diffusion model with clip latents},
  author={Ao, Tenglong and Zhang, Zeyi and Liu, Libin},
  journal={ACM Transactions on Graphics (TOG)},
  volume={42},
  number={4},
  pages={1--18},
  year={2023},
  publisher={ACM New York, NY, USA}
}

@article{hsu2021hubert,
  title={Hubert: Self-supervised speech representation learning by masked prediction of hidden units},
  author={Hsu, Wei-Ning and Bolte, Benjamin and Tsai, Yao-Hung Hubert and Lakhotia, Kushal and Salakhutdinov, Ruslan and Mohamed, Abdelrahman},
  journal={IEEE/ACM transactions on audio, speech, and language processing},
  volume={29},
  pages={3451--3460},
  year={2021},
  publisher={IEEE}
}

@inproceedings{pavlakos2019expressive,
  title={Expressive body capture: 3d hands, face, and body from a single image},
  author={Pavlakos, Georgios and Choutas, Vasileios and Ghorbani, Nima and Bolkart, Timo and Osman, Ahmed AA and Tzionas, Dimitrios and Black, Michael J},
  booktitle={Proceedings of the IEEE/CVF conference on computer vision and pattern recognition},
  pages={10975--10985},
  publisher={IEEE Computer Society},
  address={Los Alamitos, CA, USA},
  year={2019}
}

@inproceedings{li2021audio2gestures,
  title={Audio2gestures: Generating diverse gestures from speech audio with conditional variational autoencoders},
  author={Li, Jing and Kang, Di and Pei, Wenjie and Zhe, Xuefei and Zhang, Ying and He, Zhenyu and Bao, Linchao},
  booktitle={Proceedings of the IEEE/CVF International Conference on Computer Vision},
  pages={11293--11302},
  publisher={IEEE Computer Society},
  address={Los Alamitos, CA, USA},
  year={2021}
}

@article{cassell1999speech,
  title={Speech-gesture mismatches: Evidence for one underlying representation of linguistic and nonlinguistic information},
  author={Cassell, Justine and McNeill, David and McCullough, Karl-Erik},
  journal={Pragmatics \& cognition},
  volume={7},
  number={1},
  pages={1--34},
  year={1999},
  publisher={John Benjamins}
}

@article{loehr2007aspects,
  title={Aspects of rhythm in gesture and speech},
  author={Loehr, Daniel},
  journal={Gesture},
  volume={7},
  number={2},
  pages={179--214},
  year={2007},
  publisher={John Benjamins}
}

@inproceedings{kucherenko2020gesticulator,
  title={Gesticulator: A framework for semantically-aware speech-driven gesture generation},
  author={Kucherenko, Taras and Jonell, Patrik and Van Waveren, Sanne and Henter, Gustav Eje and Alexanderson, Simon and Leite, Iolanda and Kjellstr{\"o}m, Hedvig},
  booktitle={Proceedings of the 2020 international conference on multimodal interaction},
  pages={242--250},
  publisher={Association for Computing Machinery},
  address={New York, NY, USA},
  year={2020}
}

@article{xu2024chain,
  title={Chain of generation: Multi-modal gesture synthesis via cascaded conditional control},
  author={Xu, Zunnan and Zhang, Yachao and Yang, Sicheng and Li, Ronghui and Li, Xiu},
  journal={Proceedings of the AAAI Conference on Artificial Intelligence},
  volume={38},
  number={6},
  pages={6387--6395},
  year={2024}
}

@inproceedings{liu2022disco,
  title={{DisCo}: Disentangled Implicit Content and Rhythm Learning for Diverse Co-Speech Gestures Synthesis},
  author={Liu, Haiyang and Iwamoto, Naoya and Zhu, Zihao and Li, Zhengqing and Zhou, You and Bozkurt, Elif and Zheng, Bo},
  booktitle={Proceedings of the 30th ACM International Conference on Multimedia},
  pages={3764--3773},
  publisher={Association for Computing Machinery},
  address={New York, NY, USA},
  year={2022}
}

@inproceedings{chen2024enabling,
  title={Enabling synergistic full-body control in prompt-based co-speech motion generation},
  author={Chen, Bohong and Li, Yumeng and Ding, Yao-Xiang and Shao, Tianjia and Zhou, Kun},
  booktitle={Proceedings of the 32nd ACM International Conference on Multimedia},
  pages={6774--6783},
  publisher={Association for Computing Machinery},
  address={New York, NY, USA},
  year={2024}
}

@inproceedings{petrovich2021action,
  title={Action-conditioned 3D human motion synthesis with transformer VAE},
  author={Petrovich, Mathis and Black, Michael J and Varol, G{\"u}l},
  booktitle={Proceedings of the IEEE/CVF International Conference on Computer Vision},
  pages={10985--10995},
  publisher={IEEE Computer Society},
  address={Los Alamitos, CA, USA},
  year={2021}
}

@article{baevski2020wav2vec,
  title={wav2vec 2.0: A framework for self-supervised learning of speech representations},
  author={Baevski, Alexei and Zhou, Yuhao and Mohamed, Abdelrahman and Auli, Michael},
  journal={Advances in neural information processing systems},
  volume={33},
  pages={12449--12460},
  year={2020}
}

@article{henter2020moglow,
  title={Moglow: Probabilistic and controllable motion synthesis using normalising flows},
  author={Henter, Gustav Eje and Alexanderson, Simon and Beskow, Jonas},
  journal={ACM Transactions on Graphics (TOG)},
  volume={39},
  number={6},
  pages={1--14},
  year={2020},
  publisher={ACM New York, NY, USA}
}

@inproceedings{ferstl2018investigating,
  title={Investigating the use of recurrent motion modelling for speech gesture generation},
  author={Ferstl, Ylva and McDonnell, Rachel},
  booktitle={Proceedings of the 18th International Conference on Intelligent Virtual Agents},
  pages={93--98},
  publisher={Association for Computing Machinery},
  address={New York, NY, USA},
  year={2018}
}

@inproceedings{hasegawa2018evaluation,
  title={Evaluation of speech-to-gesture generation using bi-directional LSTM network},
  author={Hasegawa, Dai and Kaneko, Naoshi and Shirakawa, Shinichi and Sakuta, Hiroshi and Sumi, Kazuhiko},
  booktitle={Proceedings of the 18th International Conference on Intelligent Virtual Agents},
  pages={79--86},
  publisher={Association for Computing Machinery},
  address={New York, NY, USA},
  year={2018}
}

@article{ferstl2020adversarial,
  title={Adversarial gesture generation with realistic gesture phasing},
  author={Ferstl, Ylva and Neff, Michael and McDonnell, Rachel},
  journal={Computers \& Graphics},
  volume={89},
  pages={117--130},
  year={2020},
  publisher={Elsevier}
}

@inproceedings{takeuchi2017speech,
  title={Speech-to-gesture generation: A challenge in deep learning approach with bi-directional LSTM},
  author={Takeuchi, Kenta and Hasegawa, Dai and Shirakawa, Shinichi and Kaneko, Naoshi and Sakuta, Hiroshi and Sumi, Kazuhiko},
  booktitle={Proceedings of the 5th International Conference on Human Agent Interaction},
  pages={365--369},
  publisher={Association for Computing Machinery},
  address={New York, NY, USA},
  year={2017}
}

@inproceedings{ginosar2019learning,
  title={Learning individual styles of conversational gesture},
  author={Ginosar, Shiry and Bar, Amir and Kohavi, Gefen and Chan, Caroline and Owens, Andrew and Malik, Jitendra},
  booktitle={Proceedings of the IEEE/CVF Conference on Computer Vision and Pattern Recognition},
  pages={3497--3506},
  publisher={IEEE Computer Society},
  address={Los Alamitos, CA, USA},
  year={2019}
}

@article{alexanderson2020style,
  title={Style-Controllable Speech-Driven Gesture Synthesis Using Normalising Flows},
  author={Alexanderson, Simon and Henter, Gustav Eje and Kucherenko, Taras and Beskow, Jonas},
  journal={Computer Graphics Forum},
  volume={39},
  number={2},
  pages={487--496},
  year={2020},
  publisher={Wiley}
}

@article{wu2021modeling,
  title={Modeling the conditional distribution of co-speech upper body gesture jointly using conditional-GAN and unrolled-GAN},
  author={Wu, Bowen and Liu, Chaoran and Ishi, Carlos Toshinori and Ishiguro, Hiroshi},
  journal={Electronics},
  volume={10},
  number={3},
  pages={228},
  year={2021},
  publisher={Multidisciplinary Digital Publishing Institute}
}

@article{kucherenko2021moving,
  title={Moving fast and slow: Analysis of representations and post-processing in speech-driven automatic gesture generation},
  author={Kucherenko, Taras and Hasegawa, Dai and Kaneko, Naoshi and Henter, Gustav Eje and Kjellstr{\"o}m, Hedvig},
  journal={International Journal of Human--Computer Interaction},
  volume={37},
  number={14},
  pages={1300--1316},
  year={2021},
  publisher={Taylor \& Francis},
  doi={10.1080/10447318.2021.1883883}
}

@inproceedings{ng2021body2hands,
   title={{Body2Hands}: Learning to Infer 3D Hands from Conversational Gesture Body Dynamics},
   author={Ng, Evonne and Ginosar, Shiry and Darrell, Trevor and Joo, Hanbyul},
   booktitle={Proceedings of the IEEE/CVF Conference on Computer Vision and Pattern Recognition},
   pages={11865--11874},
   publisher={IEEE Computer Society},
   address={Los Alamitos, CA, USA},
   year={2021}
}

@inproceedings{ddpm,
    author      = {Ho, Jonathan and Jain, Ajay and Abbeel, Pieter},
    booktitle   = {Advances in Neural Information Processing Systems},
    editor      = {H. Larochelle and M. Ranzato and R. Hadsell and M.F. Balcan and H. Lin},
    pages       = {6840--6851},
    publisher   = {Curran Associates, Inc.},
    address     = {Red Hook, NY, USA},
    title       = {Denoising Diffusion Probabilistic Models},
    url         = {https://proceedings.neurips.cc/paper/2020/file/4c5bcfec8584af0d967f1ab10179ca4b-Paper.pdf},
    volume      = {33},
    year        = {2020}
}

@inproceedings{mdm,
title={Human Motion Diffusion Model},
author={Guy Tevet and Sigal Raab and Brian Gordon and Yoni Shafir and Daniel Cohen-Or and Amit Haim Bermano},
booktitle={The Eleventh International Conference on Learning Representations},
year={2023},
publisher={OpenReview.net},
address={Kigali, Rwanda},
numpages={12},
url={https://openreview.net/forum?id=SJ1kSyO2jwu}
}

@inproceedings{mld,
  title={Executing your commands via motion diffusion in latent space},
  author={Chen, Xin and Jiang, Biao and Liu, Wen and Huang, Zilong and Fu, Bin and Chen, Tao and Yu, Gang},
  booktitle={Proceedings of the IEEE/CVF Conference on Computer Vision and Pattern Recognition},
  pages={18000--18010},
  publisher={IEEE Computer Society},
  address={Los Alamitos, CA, USA},
  year={2023}
}

@inproceedings{ddim,
  author       = {Jiaming Song and
                  Chenlin Meng and
                  Stefano Ermon},
  title        = {Denoising Diffusion Implicit Models},
  booktitle    = {9th International Conference on Learning Representations, {ICLR} 2021,
                  Virtual Event, Austria, May 3-7, 2021},
  publisher    = {OpenReview.net},
  address      = {Virtual Event, Austria},
  numpages     = {22},
  year         = {2021},
  url          = {https://openreview.net/forum?id=St1giarCHLP}
}

@article{starke2022deepphase,
  title={Deepphase: Periodic autoencoders for learning motion phase manifolds},
  author={Starke, Sebastian and Mason, Ian and Komura, Taku},
  journal={ACM Transactions on Graphics (TOG)},
  volume={41},
  number={4},
  pages={1--13},
  year={2022},
  publisher={ACM New York, NY, USA}
}

@article{holden2017phase,
  title={Phase-functioned neural networks for character control},
  author={Holden, Daniel and Komura, Taku and Saito, Jun},
  journal={ACM Transactions on Graphics (TOG)},
  volume={36},
  number={4},
  pages={1--13},
  year={2017},
  publisher={ACM New York, NY, USA}
}

@inproceedings{jin2023actwishfinegrainedcontrol,
  title={Act As You Wish: Fine-Grained Control of Motion Diffusion Model with Hierarchical Semantic Graphs},
  author={Jin, Peng and Wu, Yang and Fan, Yanbo and Sun, Zhongqian and Yang, Wei and Yuan, Li},
  booktitle={Advances in Neural Information Processing Systems},
  volume={36},
  pages={15497--15518},
  publisher={Curran Associates, Inc.},
  address={Red Hook, NY, USA},
  year={2023}
}

@misc{feng2024motionwavelethumanmotionprediction,
      title={MotionWavelet: Human Motion Prediction via Wavelet Manifold Learning}, 
      author={Yuming Feng and Zhiyang Dou and Ling-Hao Chen and Yuan Liu and Tianyu Li and Jingbo Wang and Zeyu Cao and Wenping Wang and Taku Komura and Lingjie Liu},
      year={2024},
      eprint={2411.16964},
      archivePrefix={arXiv},
      primaryClass={cs.CV},
      url={https://arxiv.org/abs/2411.16964}, 
}

@inproceedings{fan2025zerozeroshotmotiongeneration,
  title={Go to Zero: Towards Zero-shot Motion Generation with Million-scale Data},
  author={Fan, Ke and Lu, Shunlin and Dai, Minyue and Yu, Runyi and Xiao, Lixing and Dou, Zhiyang and Dong, Junting and Ma, Lizhuang and Wang, Jingbo},
  booktitle={Proceedings of the IEEE/CVF International Conference on Computer Vision},
  pages={13336--13348},
  publisher={IEEE Computer Society},
  address={Los Alamitos, CA, USA},
  year={2025}
}

@article{huang2024wavedmwaveletbaseddiffusionmodels,
  title={{WaveDM}: Wavelet-Based Diffusion Models for Image Restoration},
  author={Huang, Yi and Huang, Jiancheng and Liu, Jianzhuang and Yan, Mingfu and Dong, Yu and Lv, Jiaxi and Chen, Chaoqi and Chen, Shifeng},
  journal={IEEE Transactions on Multimedia},
  volume={26},
  pages={7058--7073},
  year={2024},
  doi={10.1109/TMM.2024.3359769}
}

@inproceedings{phung2023waveletdiffusionmodelsfast,
  title={Wavelet Diffusion Models Are Fast and Scalable Image Generators},
  author={Phung, Hao and Dao, Quan and Tran, Anh},
  booktitle={Proceedings of the IEEE/CVF Conference on Computer Vision and Pattern Recognition},
  pages={10199--10208},
  publisher={IEEE Computer Society},
  address={Los Alamitos, CA, USA},
  year={2023}
}

@article{Zhang_2024,
   title={Semantic Gesticulator: Semantics-Aware Co-Speech Gesture Synthesis},
   volume={43},
   ISSN={1557-7368},
   url={http://dx.doi.org/10.1145/3658134},
   DOI={10.1145/3658134},
   number={4},
   journal={ACM Transactions on Graphics},
   publisher={Association for Computing Machinery (ACM)},
   author={Zhang, Zeyi and Ao, Tenglong and Zhang, Yuyao and Gao, Qingzhe and Lin, Chuan and Chen, Baoquan and Liu, Libin},
   year={2024},
   month={July}, pages={1--17} }

@inproceedings{chen2024tamingdiffusionprobabilisticmodels,
  title={Taming Diffusion Probabilistic Models for Character Control},
  author={Chen, Rui and Shi, Mingyi and Huang, Shaoli and Tan, Ping and Komura, Taku and Chen, Xuelin},
  booktitle={ACM SIGGRAPH 2024 Conference Papers},
  pages={1--10},
  year={2024},
  publisher={Association for Computing Machinery},
  address={New York, NY, USA},
  doi={10.1145/3641519.3657440}
}

@article{10.1145/3414685.3417838,
author = {Yoon, Youngwoo and Cha, Bok and Lee, Joo-Haeng and Jang, Minsu and Lee, Jaeyeon and Kim, Jaehong and Lee, Geehyuk},
title = {Speech gesture generation from the trimodal context of text, audio, and speaker identity},
year = {2020},
issue_date = {December 2020},
publisher = {Association for Computing Machinery},
address = {New York, NY, USA},
volume = {39},
number = {6},
issn = {0730-0301},
url = {https://doi.org/10.1145/3414685.3417838},
doi = {10.1145/3414685.3417838},
journal = {ACM Trans. Graph.},
month = nov,
articleno = {222},
numpages = {16}
}

@inproceedings{zhang2023remodiffuse,
  title={{ReMoDiffuse}: Retrieval-Augmented Motion Diffusion Model},
  author={Zhang, Mingyuan and Guo, Xinying and Pan, Liang and
          Cai, Zhongang and Hong, Fangzhou and Li, Huirong and
          Yang, Lei and Liu, Ziwei},
  booktitle={Proceedings of the IEEE/CVF International Conference
             on Computer Vision},
  pages={364--373},
  publisher={IEEE Computer Society},
  address={Los Alamitos, CA, USA},
  year={2023}
}

@inproceedings{cheng2025hologest,
  title={{HoloGest}: Decoupled Diffusion and Motion Priors for
         Generating Holisticly Expressive Co-Speech Gestures},
  author={Cheng, Yongkang and Huang, Shaoli},
  booktitle={2025 International Conference on 3D Vision},
  pages={748--757},
  year={2025},
  publisher={IEEE},
  address={Piscataway, NJ, USA}
}

@inproceedings{mughal2025raggesture,
  title={Retrieving Semantics from the Deep: An {RAG} Solution
         for Gesture Synthesis},
  author={Mughal, M. Hamza and Dabral, Rishabh and
          Scholman, Merel C. J. and Demberg, Vera and
          Theobalt, Christian},
  booktitle={Proceedings of the IEEE/CVF Conference on
             Computer Vision and Pattern Recognition},
  pages={16578--16588},
  publisher={IEEE Computer Society},
  address={Los Alamitos, CA, USA},
  year={2025}
}

@InProceedings{Lugmayr_2022_CVPR,
    author    = {Lugmayr, Andreas and Danelljan, Martin and Romero, Andres and Yu, Fisher and Timofte, Radu and Van Gool, Luc},
    title     = {RePaint: Inpainting Using Denoising Diffusion Probabilistic Models},
    booktitle = {Proceedings of the IEEE/CVF Conference on Computer Vision and Pattern Recognition (CVPR)},
    month     = {June},
    year      = {2022},
    pages     = {11461--11471},
    publisher = {IEEE Computer Society},
    address   = {Los Alamitos, CA, USA}
}

\end{document}